\documentclass[aps,showpacs,prb,twocolumn,amsmath,amssymb,superscriptaddress]{revtex4}

\usepackage{epsfig}

\begin{document}

\title{Junctions of Spin-Incoherent Luttinger Liquids with Ferromagnets and Superconductors}

\author{Dagim Tilahun}
\affiliation{Department of Physics, The University of Texas at Austin, Austin, TX 78712  USA} \affiliation{Kavli Institute for Theoretical
Physics, University of California, Santa Barbara, CA 93106 USA}
\author{Gregory A. Fiete}
\affiliation{Department of Physics, The University of Texas at Austin, Austin, TX 78712  USA}
\affiliation{Department of Physics,
California Institute of Technology, MC 114-36, Pasadena, CA 91125 USA}

\date{\today}

\begin{abstract}

We discuss the properties of a strongly interacting spin-charge separated one dimensional system coupled to ferromagnets and/or superconductors.
Our results are valid for arbitrary temperatures with respect to the spin energy, but require temperature be small compared to the charge
energy.  We focus mainly on the spin-incoherent regime where temperature is large compared to the spin energy, but small compared to the charge
energy.  In the case of a ferromagnet we study spin pumping and the renormalized dynamics of a precessing magnetic order parameter.  We find the
interaction-dependent temperature dependence of the spin pumping can be qualitatively different in the spin-incoherent regime from the Luttinger
liquid regime, allowing an identification of the former. Likewise, the temperature dependence of the renormlized magnetization dynamics can be
used to identify spin-incoherent physics.  For the case of a spin-incoherent Luttinger liquid coupled to two superconductors, we compute the ac
and dc Josephson current for a wire geometry in the 
limit of tunnel coupled superconductors. Both the ac and dc response contain ``smoking gun" signatures that can be used to identify
spin-incoherent behavior.
Experimental requirements for the observation of these effects are laid out.

\end{abstract}



\maketitle

\section{Introduction}

An exotic and distinctive feature of interacting one-dimensional systems is the phenomenon of spin-charge separation where the elementary
excitations of the fermionic system are decoupled charge and spin bosonic modes that propagate at different
velocities.\cite{Giamarchi,Gogolin,Voit:rpp95}  At low energies, Luttinger liquid theory\cite{haldane81}  has been quite successful at
describing the properties of one-dimensional systems such as quantum wires\cite{yaro02, Auslaender:sci05} and
nanotubes.\cite{Bockrath:nat99,Yao:nat99} One of the elegant features of the theory is the ease with which one can obtain exact results, thanks
to the quadratic form of the Hamiltonian describing the low energy properties,
\begin{equation}
H_{1D}  = \sum_{i}v_i\int \frac{dx}{2\pi} \left[\frac{1}{K_i}(\partial_x \theta_i(x))^2 + K_i(\partial_x \phi_i(x))^2\right],
\label{1DHamltn}
\end{equation}
where $i=\rho,\sigma$ stands for the charge and spin sectors and $\theta_i$ and $\phi_i$ are bosonic fields satisfying
$[\theta_i(x),\phi_j(x')]=-i\frac{\pi}{2} \delta_{ij}{\rm sgn}(x-x')$, and representing charge/spin density fluctuations and charge/spin current
density fluctuations, respectively, and $v_i$ are the collective mode velocities. The parameters $K_i$ describe the interactions in the 1-d
system. For repulsive interactions $K_\rho<1$, and for SU(2) symmetry in the spin sector $K_\sigma=1$. Generally, the theory is applied under the
assumption that both the spin and charge modes operate at low energy. But at low densities, $n$, where $r_s=(2na_B)^{-1}\gg 1$ (with $a_B$ the
Bohr radius for the material), strong interactions open up a wide window between the characteristic energies of the charge and the spin sectors
by suppressing the spin exchange energy $E_\sigma$ while enhancing the charge energy $E_\rho$. When the temperature $T$ is raised into this window
of energy, $E_\sigma \ll k_B T \ll E_\rho$ (where $k_B$ Boltzmann's constant), the spin sector consists of thermally randomized spins that are
no longer described by (\ref{1DHamltn}) while the charge sector effectively becomes a spinless Luttinger Liquid (LL). A one-dimensional system
in this regime is known as a spin-incoherent Luttinger liquid (SILL).\cite{Fiete:rmp07}

Several properties of the SILL have already been established theoretically,\cite{Fiete:rmp07} but the experimental effort has been slowed due to
the difficulty of reaching the aforementioned window of energy. Nevertheless, there are experimental indications of this regime in high quality
quantum wires grown with the cleaved-edge technique,\cite{Steinberg:prb06} in recent split-gate devices,\cite{Hew:prl08} and possibly even in
low-density carbon nanotubes.\cite{Deshpande:np08} Given the recent progress in manufacturing hybrid devices consisting of superconductors (SC)
and ferromagnets (FM), and one-dimensional systems such as nanotubes, we propose here that the properties of the SILL be experimentally
investigated by studying junctions of superconductors and ferromagnets with a SILL. As we show below, these SILL hybrid systems lead to
distinctive behaviors (relative to a LL or non-interacting 1-d system) that will allow the SILL to be identified in various types of transport
measurements, and in the dynamics of a coupled ferromagnet.

In early work on transport in the spin-incoherent regime, Matveev found that a SILL adiabatically connected to non-interacting leads resulted in
a universal reduction of the conductance of a single-channel quantum wire from $2 e^2/h$ to $e^2/h$.\cite{Matveev:prl04,Matveev:prb04}  This
effect has been ascribed to the deactivation of the spin channel, whereby the spin excitations are reflected while the charge ones propagate
freely.  In a recent related work, the authors studied the problem of a SILL adiabatically connected to two superconductors.  It was found that
the critical Josephson current through such a SC-SILL-SC systems suffers the same fate: the critical value of the Josephson current is halved
relative to the LL case.\cite{Tilahun:prb08}   These universal results follow from the assumption of adiabaticity of the contacts between the
SILL and the leads/SC, which may not be satisfied in some experimental situations.

In this work we study the transport properties of a SILL in the opposite limit where it is contacted via tunnel junctions to a superconductor
and/or a ferromagnet.  We will discuss several scenarios that may be implemented experimentally with existing technologies.  The central aim of
this work is to contribute to our fundamental understanding of the SILL by addressing its properties in hybrid structures, and to add to the
arsenal of experiments that can be carried out to probe it and infer its existence.

Our main results are the following.  For the case of a FM tunnel coupled to a SILL, we compute the spin current pumped into the SILL due to a
time-dependent magnetization in the ferromagnet.  The algebraic form of the expression for the pumped spin current is the same as that for
non-interacting fermions, or a Luttinger liquid, but the coefficients appearing have a temperature dependence characteristic of the SILL.  We
argue the spin transport is diffusive and under this assumption compute the renormalization (due to the pumped spin current and its back-flow)
of the Gilbert damping $\alpha$ of the magnetization motion in the FM.  This also has a characteristic temperature dependence in the
spin-incoherent regime. For the case of a SC tunnel coupled to a SILL, we evaluate the ac and dc Josephson current in a wire that can be either
side-coupled (bulk-coupled) or end-coupled to the SC.  We find the dc Josephson current suffers an exponential decay in real space due to the incoherence of
the spin sector (in contrast to the inverse length suppression that was found for adiabatically contacted SCs\cite{Tilahun:prb08}). Finally, we show that the ac Josephson effect also contains important information about
spin-incoherent effects in its voltage and temperature dependence.

This paper is organized as follows. In Sec.~\ref{sec:bosonization} we review bosonization for a strongly interacting electron system for energy
scales small compared to the charge energy, but arbitrary compare to the spin energy. In 
Sec.~\ref{sec:FM} we study spin pumping and
magnetization dynamics in the FM-SILL system.  In Sec.~\ref{sec:SC-SILL}  we study ac and dc Josephson effects in the SC-SILL-SC system.
Finally in Sec.~\ref{sec:discussion} we discuss prospects for experimental realizations of the physics we discussed and directions for future
work. Some technical details and results appear in various appendicies.

\section{Review of Bosonization for Strongly Interacting Electrons}
\label{sec:bosonization}

In this paper we will be studying a strongly interacting one dimensional electron system coupled to either a ferromagnet or a superconductor.
Strong interactions imply\cite{Fiete:rmp07} that $E_\sigma \ll E_\rho$ and therefore it is possible to be in a regime of temperatures that may
not be small compared to the spin energy $E_\sigma$, but are still small compared to the charge energy $E_\rho$.  This means that we are no
longer free to use standard bosonization procedures\cite{Giamarchi,Gogolin,Voit:rpp95} for the electron operator because these formulas rely on
the assumption that the relevant energy scales on which the system is probed are small compared to both spin and charge energies.

For a strongly interacting system we require a formalism that applies to all energy scales relative to the spin energy (and in particular to
$k_B T \gtrsim E_\sigma$), but only low energy scales to relative to the charge energy.  Such a formalism was recently developed by Matveev,
Furusaki, and Glazman,\cite{Matveev:prb07} and applied to the evaluation of the spectral function and tunneling into a strongly interacting
electron system at arbitrary temperatures relative to the spin energy.\cite{Matveev:prl07}  In this section we review the essential elements of
their work and show how to represent the electron operator in a strongly interacting one dimensional electron system.   The general expressions
for electron correlation functions obtained in this way provide a launching point for numerical studies in the intermediate (with respect to
spin) temperature regime $k_B T \approx E_\sigma$, which is not well handled by existing analytical methods.

\subsection{Basic assumptions and general expressions}

As is typical of interacting one dimensional electron systems, we assume the Hamiltonian in the strongly interacting case is spin-charge
separated,\cite{Matveev:prb04, Fiete:rmp07} $H=H_\rho+H_\sigma$.  Here, $H_\rho$ is identical to that of the LL, and is given by
Eq.~\eqref{1DHamltn}.  On the other hand, the spin Hamiltonian at arbitrary temperatures is to a very good approximation given by a nearest
neighbor antiferromagnetic Heisenberg spin chain,\cite{Matveev:prb04,Klironomos:prb05,Fogler_exch:prb05}
\begin{equation}
\label{eq:S_chain}
H_\sigma = \sum_l J {\vec S}_l\cdot {\vec S}_{l+1},
\end{equation}
where evidently the spin energy is set by $J$: $E_\sigma=J$.  The basic idea is to represent the electron operator as a product of operators
that describe the holons $\Psi(x)$ (spinless fermions that naturally arise in the context of strongly interacting fermions and the
spin-incoherent regime\cite{Fiete_2:prb05,Fiete:prl04}) and the spin degrees of freedom ${\vec S_l}$.  The holon operators (denoted by
$\Psi^\dagger,\Psi$) by construction satisfy the equation
\begin{equation}
\label{eq:Psi}
\Psi^\dagger(x)\Psi(x)=\psi_\uparrow^\dagger(x)\psi_\uparrow(x)+\psi_\downarrow^\dagger(x)\psi_\downarrow(x),
\end{equation}
where $\psi_s$ is the electron annihilation operator for electrons of spin projection $s$, and $\psi^\dagger_s$ the corresponding electron
creation operator.

The issue of how to bosonize the electron operator for a strongly interacting system earlier arose in the context of the large $U$ limit of the
one dimensional Hubbard model.  Penc {\it et al.}\cite{Penc:prl95} wrote the electron creation operator as
\begin{equation}
\label{eq:Penc_bos}
\psi_s^\dagger(0)=Z^\dagger_{0,s} \Psi^\dagger(0),
\end{equation}
where $Z^\dagger_{0,s}$ creates a site on the spin chain \eqref{eq:S_chain} with spin projection $s$.  The expression \eqref{eq:Penc_bos} can be
physically motivated as follows.  From \eqref{eq:Psi} it is clear that the creation of an electron is also accompanied by the creation of a
holon.  However, electrons also carry spin so there must be a component of the electron operator that also creates spin.  This is accomplished
by $Z^\dagger_{0,s}$.  In general, one has $Z^\dagger_{l,s}$ as the object that adds a new site to the spin chain between $l-1$ and $l$.  While
this appears physically intuitive, the expression suffers from the drawback that it does not naturally account for the variation of electron
density with position in a real electron gas.\cite{Penc:prl96}  Matveev, Furusaki, and Glazman showed\cite{Matveev:prb07} that a remedy for this
issue is to define the position at which the spin site is added to the chain \eqref{eq:S_chain} in terms of the number of holons to the left of the site,
\begin{equation}
\label{eq:l_x}
l(x)=\int_{-\infty}^x \Psi^\dagger(y)\Psi(y) dy.
\end{equation}
In terms of \eqref{eq:l_x} the electron creation and annihilation operators are defined as
\begin{eqnarray}
\label{eq:psi+_x}
\psi_s^\dagger(x)&=&Z^\dagger_{l(x),s} \Psi^\dagger(x),\\
\label{eq:psi_x}
\psi_s(x)&=&\Psi(x) Z_{l(x),s}.
\end{eqnarray}
The operators given above explicitly account for the fact that the spins are attached to electrons, and the formulas are valid at all energy
scales. It is perhaps worth noting that even though the Hamiltonian is spin-charge separated, the electron operators are not written as a
product of a spin piece and a charge piece because the ``spin" pieces $Z_{l(x),s}$ also depend on the electron density via \eqref{eq:l_x}.

\subsection{Bosonizing the holon operators}

In writing Eqs.\eqref{eq:psi+_x},\eqref{eq:psi_x} no assumptions have been made about the energy scale relative to the spin and charge energies.
We now restrict our considerations to energies small compared to $E_\rho$, but arbitrary with respect to $E_\sigma$.  In this case, we are free
to bosonize the holon sector.  The Hamiltonian for the holons must then necessarily take the low energy form\cite{Fiete_2:prb05}
\begin{equation}
H_{\rho}  = v_\rho \int \frac{dx}{2\pi} \left[\frac{1}{K}(\partial_x \theta(x))^2 + K(\partial_x \phi(x))^2\right],
\label{eq:H_holon}
\end{equation}
where the interaction parameter $K$ of the holon is related to the interaction parameter of the charge sector as\cite{Fiete_2:prb05}
$K=2K_\rho$, and the spinless fields $\theta$ and $\phi$ can be related to the holon density as\cite{Fiete:prb05,Fiete:prl04}
\begin{equation}
\label{eq:den_rel}
\Psi^\dagger(x)\Psi(x)=\frac{1}{\pi}[k_F^h + \partial_x \theta(x)],
\end{equation}
where the holon Fermi wave vector is twice the electron Fermi wavevector\cite{Fiete:prl04} $k_F^h=2k_F$.  The bosonic fields satisfy the
commutation relations $[\theta(x),\partial_y\phi(y)]=i\pi\delta(x-y)$.

Since we are interested in low energies with respect to the charge energy, the electron operator may be expanded about the two holon Fermi
points at $\pm k_F^h$,
\begin{equation}
\label{eq:Psi_RL}
\Psi(x)=\Psi_R(x)+\Psi_L(x),
\end{equation}
where $\Psi_R(x)$ destroys an holon near the right Fermi point and $\Psi_L(x)$ destroys an electron near the left Fermi point. The left and
right holon operaters are bosonized as
\begin{equation}
\label{eq:Psi_RL_fields}
\Psi_{R,L}(x)=\frac{1}{\sqrt{ 2\pi \alpha_c}} e^{-i\phi(x)}e^{\pm i[k_F^hx+\theta(x)]},
\end{equation}
where $\alpha_c$ is a short distance cut-off of order the interparticle spacing $a$.  Combining the results of \eqref{eq:l_x}, \eqref{eq:psi_x},
\eqref{eq:den_rel}, \eqref{eq:Psi_RL} and \eqref{eq:Psi_RL_fields} one obtains the bosonized form of the electron annihilation operator for spin
$s$
\begin{eqnarray}
\label{eq:psi_bos}
\psi_s(x)=\frac{e^{-i\phi(x)}}{\sqrt{2 \pi \alpha_c}}\left(e^{i[k_F^hx+\theta(x)]}+e^{-i[k_F^hx+\theta(x)]}\right)\nonumber \\
\times Z_{l,s}\bigg |_{l=\frac{1}{\pi}[k_F^hx+\theta(x)]},
\end{eqnarray}
and an analogous expression for the electron creation operator $\psi^\dagger_s(x)$.  Expression \eqref{eq:psi_bos}, however, it not quite
complete as it does not account for the discreteness of the charge of the electron.  This can be accomplished by interpreting
\begin{equation}
\label{eq:Z_proper}
Z_{l,s}\bigg |_{l=\frac{1}{\pi}[k_F^hx+\theta(x)]} \to \sum_l Z_{l,s}\delta\left({1 \over \pi}[k_F^hx+\theta(x)]-l\right),
\end{equation}
after which the full electron annihilation operator (include both left and right moving parts) becomes\cite{Matveev:prb07}
\begin{equation}
\label{eq:psi_final_bos}
\psi_s(x)=\frac{e^{-i\phi(x)}}{\sqrt{2\pi\alpha_c}} \int_{-\infty}^\infty \frac{dq}{2\pi} z_s(q) e^{i(1+{q \over \pi})[k_F^hx+\theta(x)]},
\end{equation}
where
\begin{equation}
\label{eq:little_z}
z_s(q)=\sum_{l=-\infty}^\infty Z_{l,s}e^{-iql}.
\end{equation}
The expression for the electron annihilation operator \eqref{eq:psi_final_bos} is the key result of the this section, obtained earlier by
Matveev, Furusaki, and Glazman, who also showed that in the limit of small energies compared to $E_\sigma$ the expression correctly recovers the
standard LL formulas for the electron annihilation operator.\cite{Matveev:prb07}  With \eqref{eq:psi_final_bos}  correlation functions involving
electron operators can be expressed in terms of the correlation functions of the holon and spin sectors at arbitrary temperatures with respect
to $E_\sigma$, but small energies compared to $E_\rho$. See Appendix~\ref{app:spin_current}  and Appendix~\ref{app:Andreev_TDOS} for examples of
the evaluation of the single particle Green's function near different types of boundaries  using the bosonization formula
\eqref{eq:psi_final_bos} and the boundary conditions to be discussed next.

\subsection{Open and Andreev boundary conditions}

In this work we will be primarily interested in two types of boundary conditions on the electron operators: (1) open or ``hard wall" boundary
conditions appropriate for tunnel junctions and (2) Andreev boundary conditions appropriate for adiabatically contacted (no electron
backscattering) superconductors.  Other, more general, ``intermediate" boundary conditions are possible though less generic.\cite{Affleck:prb00}
To help orient the reader, in each of the two cases above we will briefly summarize the result appropriate for the LL regime for the purpose of
drawing contrast with the strongly interacting, finite temperature SILL regime.  Our discussion of the boundary conditions extends the results
of Ref.~[\onlinecite{Matveev:prb07}].

\subsubsection{Open boundary conditions}
\label{subsec:open}

Open boundary conditions, or ``hard wall" boundary conditions result in electron waves that are perfectly reflected at the boundary. This
implies there is no charge current or spin current through the boundary.  For concreteness, let us assume that our boundary is located at $x=0$
with the interacting one dimensional system living on $x>0$.  Then an electron traveling to the left with spin $s$ will be reflected to a right
moving electron with the same spin $s$
\begin{equation}
\label{eq:bc_open}
\psi_{L,s}(0)=e^{-i \eta}\psi_{R,s}(0),
\end{equation}
with  a phase shift $\eta$ that depends on details of the boundary scattering potential.\cite{Affleck:prb00}
From the standard LL bozonization formulas (in our convention)
\begin{equation}
\label{eq:standard_bos}
\psi_{R/L,s}(x)=\frac{1}{\sqrt{2\pi \alpha_c}} e^{-i\phi_s(x)}e^{\pm i [k_F x +\theta_s(x)]},
\end{equation}
 we see that \eqref{eq:bc_open} implies that $\theta_s(0)={\rm const}$, and therefore also that $\theta_\rho(0)={\rm const}$ and
 $\theta_\sigma(0)={\rm const}$. In computing correlation functions at the boundary, this means that we must take $\theta_\rho(0)$ and
 $\theta_\sigma(0)$ to be non-fluctuating quantities, allowing only $\phi_\rho(0)$ and $\phi_\sigma(0)$ to fluctuate.

In the strongly interacting case, the reflection condition \eqref{eq:bc_open} implies that $\theta(0)={\rm const}$, where we have used
\eqref{eq:psi_bos}.  Since  the spin site $l$ is related to the $\theta$ field via \eqref{eq:den_rel} and \eqref{eq:l_x} we have $l\equiv 0$ at
$x=0$.  This implies any correlation function involving electron operators evaluated at the boundary will depend only on $Z_{l=0,s}$ or its
Hermitian conjugate, and the fluctuations of the field $\phi$ at $x=0$.  We will apply these boundary conditions in a calculation of the single
particle Green's function in Appendix~\ref{app:spin_current}.

For completeness, below we give the expansions of the bosonic fields for a system of finite length $L$ with open boundary conditions at $x=0$
and $x=L$:\cite{Kane:prl97,Fiete:prb05,Fabrizio:prb95}
\begin{eqnarray}
\label{eq:open_expansions}
\theta(x)&=&i\sum_{m=1}^\infty \sqrt{\frac{2K_\rho}{m}}\sin\left(\frac{m \pi x}{L}\right)(b_m -b^\dagger_m)+\theta^0(x), \nonumber \\
\phi(x)&=&\sum_{m=1}^\infty \sqrt{\frac{1}{2K_\rho m}}\cos\left(\frac{m \pi x}{L}\right)(b_m +b^\dagger_m)+\Phi,
\end{eqnarray}
where $\theta^0(x)=\frac{\pi x}{L} N$, $[b_n,b^\dagger_m]=\delta_{nm}$, and $[\Phi,N]=i$. For a semi-infinite system, we take $L\to \infty$ and
the discrete sums over $m$ become integrals over momentum $q_m=m\pi/L$.

\subsubsection{Andreev boundary conditions}
\label{subsec:Andreev}

In a certain sense the Andreev limit is the opposite limit of ``hard wall" scattering from a SC-M, or SC-LL, or SC-SILL interface. Whereas open
boundary conditions imply an incident electron is perfectly reflected, Andreev boundary conditions imply that an electron is perfectly absorbed
by the SC (with a concomitant reflected hole of the opposite spin).\cite{Andreev:jetp64}  However, compared to the open boundary conditions
\eqref{eq:bc_open} the Andreev boundary conditions are more subtle as they involve an energy scale, the superconducting gap $\Delta$, that under
many circumstances cannot be taken to be infinitely large relative to the energy scales of interest (such as $k_B T$, or an applied voltage) in
the interacting one-dimensional system.  One of the most important consequences of a finite $\Delta$ is an energy-dependent reflection
coefficient, which ultimately leads to the proximity effect in the normal material.\cite{deGennes:rmp64} In the context of interacting one
dimensional systems, Andreev boundary conditions have been discussed by a number of
authors.\cite{Affleck:prb00,Maslov:prb96,Vishveshwara:prb02,Takane_Koyama:jpsj97,Takane:jpsj96,Titov:prl06,Lee:prl03}  The conclusion of these
works is that when a left moving electron with spin $s$ is reflected as a right moving hole with the opposite spin there is an energy dependent
phase shift $e^{i q \xi}$ (proportional to the momentum difference $q$ with respect to the Fermi point)\cite{Takane:jpsj96} that multiplies a
factor\cite{Maslov:prb96} $e^{i\chi}$ that encodes the phase $\chi$ of the superconducting order parameter  (assumed non-zero for
$x<0$),\cite{Affleck:prb00}
\begin{equation}
\label{eq:bc_Andreev}
\psi_{L,s}(0)=(-1)^{f(s)} ie^{i\chi} e^{i q \xi}\psi^\dagger_{R,-s}(0),
\end{equation}
where $\xi \propto 1/\Delta$ is the superconducting coherence length.  The function $f(s)=0$ for $s=\uparrow$ and $f(s)=1$ for $s=\downarrow$.
The boundary conditions \eqref{eq:bc_Andreev} are valid only at energy scales much smaller than $\Delta$, which we will assume throughout this
work.

Applying the boundary conditions \eqref{eq:bc_Andreev} to the LL case (with bosonized electron operator below \eqref{eq:bc_open}) gives
$\theta_\sigma \propto  \theta_s-\theta_{-s}={\rm const}$ and $ \phi_\rho \propto \phi_s+\phi_{-s}= {\rm const}$.  Therefore, we find that much
like the situation of perfect reflection there is no spin-current through the interface, but there is is a net charge current.  Moreover,
analysis of singlet superconductivity and spin density wave correlation functions using \eqref{eq:bc_Andreev} in the LL regime shows that there
are suppressed spin fluctuations near (distances less than $\xi$) the interface.\cite{Takane:jpsj96}  It is perhaps worth noting that if the
SC-LL interface has a very weak electron backscattering, the interactions in the LL tend to renormalize the interface
scattering.\cite{Takane_Koyama:jpsj97,Titov:prl06,Affleck:prb00}

For strongly interacting electrons the boundary condition  \eqref{eq:bc_Andreev} implies $\phi(0)={\rm const}$ and $\sum_l
Z_{l,s}\delta\left(\theta(0)/\pi-l\right)=\sum_l Z^\dagger_{l,-s}\delta\left(\theta(0)/\pi-l\right) \implies Z_{l,s}=Z^\dagger_{l,-s}$ at the
interface, where we have again used \eqref{eq:psi_bos} and also \eqref{eq:Z_proper}.  We show in Appendix~\ref{app:Andreev_TDOS} that we recover
our earlier results\cite{Tilahun:prb08} for SC-SILL junctions in the Andreev limit using this formalism.

Finally, for completeness and for unification of our notation, below we give the expansions of the bosonic
fields\cite{Maslov:prb96,Tilahun:prb08,Fazio:prb96} for Andreev boundary conditions on a system of length $L$ (see Fig.~\ref{fig:SC_tunnel})
with superconducting order parameter phase difference $\chi=\chi_1-\chi_2$:
\begin{eqnarray}
\label{eq:Andreev_expansions}
\theta(x)&=&\sum_{m=1}^\infty \sqrt{\frac{2K_\rho}{m}}\cos\left[\frac{m \pi }{L}\left(x+{\xi \over 2}\right)\right](b_m +b^\dagger_m)+\theta^0,
\nonumber  \\
\phi(x)&=&i\sum_{m=1}^\infty \sqrt{\frac{1}{2K_\rho m}}\sin\left[\frac{m \pi }{L}\left(x+{\xi \over 2}\right)\right](b_m -b^\dagger_m)\nonumber \\
&& \hspace{5cm}+\Phi(x),
\end{eqnarray}
where $\Phi(x)=\pi\left (J'+\frac{\chi}{\pi}\right)\frac{x}{L}$, $[b_n,b^\dagger_m]=\delta_{nm}$, and $[\theta^0,J']=i$. The topological number
$J'=(N_\uparrow+N_\downarrow)/2+1$ and the total spin of the system $M\equiv (N_\uparrow-N_\downarrow)/2$ must satisfy the constraint
$J'+M=$even.\cite{Maslov:prb96}  We have also included the proximity effects\cite{Takane:jpsj96} via the length $\xi$ in the field expansions
that appeared earlier in the general expression \eqref{eq:bc_Andreev} for the boundary conditions appropriate to Andreev
reflection.\cite{Takane_remark}

Note that compared to the field expansions for open boundary conditions \eqref{eq:open_expansions}, the expansions for Andreev boundary
conditions \eqref{eq:Andreev_expansions} have $\theta$ and $\phi$ ``switched".  This can be understood in simple physical terms: $\theta$ and
$\phi$ are conjugate fields so if one is a constant, the other is strongly fluctuating.  Thus, the ``switching" of the fields comes from
``opposite" nature of the two boundary conditions in the charge sector. For open boundary conditions there is no charge current through the
boundary, while for Andreev boundary conditions the charge current on either side of the boundary is unchanged by its presence  because there is
no electron backscattering there.  Had we been concerned with the bosonization of the spin sector, we would have found that due to the absence
of spin current through the boundary that sector would have expansions appropriate for open boundary conditions similar to
\eqref{eq:open_expansions}.\cite{Maslov:prb96,Tilahun:prb08,Fazio:prb96}

Having spent the time to develop the formalism used in our calculations, we now turn our attention squarely to the physics of hybrid junctions
involving spin-incoherent Luttinger liquids.  We first discuss a junction consisting of a ferromagnet tunnel coupled to a spin-incoherent
Luttinger liquid.

\section{FM-SILL Tunnel Junctions}
\label{sec:FM}

\begin{figure}[h]
\includegraphics[width=.65\linewidth,clip=]{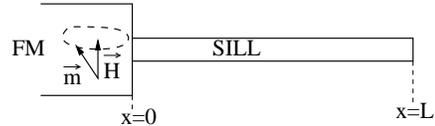}
 \caption{\label{fig:FM} Schematic of the model we study. A ferromagnet is tunnel coupled at $x=0$ to a spin-incoherent Luttinger liquid of length $L.$
 As the magnetization vector ${\vec m}$ precesses about an effective magnetic field ${\vec H}$ spin is pumped into the SILL.  Spin accumulation in the
 SILL will lead to renormalization of the magnetization dynamics.}
 \end{figure}

\subsection{Spin pumping}
We consider the set-up shown schematically in Fig.\ref{fig:FM} whereby a ferromagnet is coupled via a tunnel juntion to a spin-incoherent
Luttinger liquid. A similar situation was considered by Bena and Balents in the context of a FM-LL junction.\cite{Bena:prb04} They found that
when the magnetization $\vec{m}$ of the FM acquires a time dependence (perhaps by the application of external fields), spin current is pumped
into the adjoined LL at the rate
\begin{equation}
\langle \vec I(t)\rangle  = -A_1\vec{m}\times\frac{d\vec{m}}{dt}+A_2\frac{d\vec{m}}{dt},
\label{spnexpr}
\end{equation}
where $A_1$ and $A_2$ are temperature dependent parameters depending on the interface tunneling and interactions in the LL. Below we show that
an identical expression is obtained for a SILL and we derive explicit expressions for $A_1$ and $A_2$ in this case.  We schematically indicate
how the temperature dependence crosses over from the LL to the SILL case.  We note that \eqref{spnexpr} was earlier derived for a
non-interacting metal attached to a FM.  In this case, $A_1$ and $A_2$ are temperature independent to lowest
order.\cite{Tserkovnyak:prb02,Tserkovnyak:prl02,Tserkovnyak:rmp05}  Thus, the temperature dependence of $A_1$ and $A_2$ encodes information
about the interactions and  may also indicate whether the attached 1-d system is in the LL or SILL regime.  We now turn to a derivation of these
results.

We assume that the ferromagnet is itinerant, such as Fe or Co, and can be described by a Stoner-type model in which there is a different density
of states for spin-up and spin-down electrons.  The different density of states can be absorbed into distinct effective tunneling matrix
elements, $t_s$, for spin-up and spin-down electrons.\cite{Brataas:prl00,Balents:prb01}  This also implies that the local
action\cite{local_S_remark} describing the spin up and spin down electrons are identical to that of non-interacting electrons
\begin{equation}
\label{eq:S_FM}
S_{\rm FM}=\frac{1}{\beta}\sum_{\omega_n} \sum_{s=\uparrow,\downarrow} \frac{ |\omega_n|}{2\pi} |\phi^s_m(\omega_n)|^2,
\end{equation}
where $\beta$ is the inverse temperature and $\phi^s_m(\omega_n)$ are the bosonic fields describing local fluctuations at the tunneling point,
$x=0$, of spin up and spin down electrons in the FM.  For the SILL we have the following spin-charge separated form of the local action
\begin{equation}
\label{eq:S_SILL}
S_{\rm SILL}=\frac{1}{\beta} \sum_{\omega_n}  \frac{K_\rho |\omega_n|}{2\pi} |\phi^\rho(\omega_n)|^2+S_{\rm SILL}^\sigma,
\end{equation}
where $0<K_\rho<1$ is the Luttinger parameter of the charge sector of the SILL, $\phi^\rho=(\phi^\uparrow+\phi^\downarrow)/\sqrt{2}$ is the
bosonic charge field, and $S_{\rm SILL}^\sigma$ is the action for the spin sector which will actually play no role in the evaluation of
correlation functions in the spin-incoherent regime as it effectively drops out leading to ``super universal spin physics".\cite{Fiete:rmp07}

For an arbitrary spin quantization axis relative to the magnetization ${\vec m}$ it is useful to define
\begin{equation}
\hat u_\pm=(1\pm \hat m \cdot \vec \sigma)/2,
\end{equation}
which projects the spin quantization axis onto the magnetization direction $\hat m$. Here $u_\pm$ is a $2\times 2$ matrix and $\vec \sigma$ are
the Pauli spin matricies.  The tunneling Hamiltonian then takes the form\cite{Balents:prb01,Balents:prl00}
\begin{equation}
H^{\rm gen}_{\rm tun}=F^\dagger W\psi+\psi^\dagger W^\dagger F,
\label{eq:tun_gen}
\end{equation}
where $F$ ($F^\dagger$) annihilates/creates an electron in the FM and $\psi$ ($\psi^\dagger)$ annihilates/creates an electron in the SILL. The
tunneling is assumed to occur at $x=0$ as shown in Fig.\ref{fig:FM}.  The $2\times2$ tunneling matrix $W$ is then
\begin{equation}
W=\sum_{s=\pm} t_s \hat u_s.
\end{equation}
The tunneling Hamiltonian \eqref{eq:tun_gen} can be expressed more explicitly as
\begin{equation}
H_{\rm tun}  = \sum_{s} u_1 F^\dagger_s \psi_s + \vec{m}(t)\cdot \sum_{s,s'}u_2F^\dagger_s\frac{\vec{\sigma}_{s,s'}}{2}\psi_{s'} + h.c.,\label{tunHam}
\end{equation}
where $\vec{m}(t)$ is the time-dependent magnetization of the ferromagnet, and $u_1=(t_+ + t_-)/2$ and $u_2=(t_+-t_-)/|\vec m|$. (From here
onwards we will assume $|\vec m|=1$.) The spin current operator is obtained from the relation $ \vec{I}(t) = \frac{d\vec{M}}{dt} =
-\frac{i}{\hbar}[\vec{M},H_{\rm tun}]$, where $\vec{M} = \frac{1}{2}\psi_s^\dagger\vec{\sigma}_{ss'}\psi_{s'}$ is the spin density in the SILL
at the boundary and summation over repeated spin indices is understood. It follows then that the spin current operator is given by
\begin{equation}
\vec{I}(t)  = \frac{iu_1}{2}F^\dagger_s\vec{\sigma}_{ss'}\psi_{s'}+\frac{iu_2}{4}\vec{m}F^\dagger_s\psi_s +
\frac{u_2}{4}F^\dagger_s\vec{m}\times\vec{\sigma}_{ss'}\psi_{s'}+h.c.\label{spncurnt}
\end{equation}
The spin current, $\langle \vec{I}(t) \rangle =-\frac{i}{\hbar}\int dt' \Theta(t-t')\langle [I(t),H_{\rm tun}(t')] \rangle$, is
obtained\cite{Bena:prb04} from second-order time dependent perturbation theory which gives
\begin{eqnarray}
\label{spncurntfnl}
\langle\vec{I}(t)\rangle  = {\vec m}(t)\text{Im}
(u_2^*u_1)\text{Re}[C(0)]-\int \frac{d \omega}{h} C(\omega)e^{-i\omega t}\times\nonumber\\\left(\text{Im}
(u_2^*u_1)\vec{m}(\omega)+\frac{|u_2|^2}{2}\vec{m}(\omega)\times\vec{m}(t)\right),
\end{eqnarray}
where $C(\omega)$ is the Fourier transform of the retarded Greens function $C(t-t') =
-i\Theta(t-t')\langle[F^\dagger_s(t)\psi_s(t),\psi^\dagger_s(t')F_s(t')]\rangle$. If we assume the typical frequencies of the magnetization
precession are in the GHz range\cite{Tserkovnyak:rmp05} (which corresponds to energies of roughly 100 mK) or smaller, then this is a small
energy scale in the problem and we may safely expand $C(\omega)$ for small $\omega$. It can be easily checked that the zero frequency terms
cancel exactly, leaving only the contributions linear in $\omega$, provided we drop terms proportional to $\omega^2$ and all higher powers.
Upon integration over frequency, the linear frequency terms are converted to time derivatives yielding expression \eqref{spnexpr} where the
coefficients $A_1$ and $A_2$ are given by
\begin{equation}
\label{eq:A_gen}
A_1=-\frac{i}{h} C'(0) \frac{|u_2|^2}{2},\;\;  A_2 =-\frac{i}{h} C'(0) {\rm Im}(u_2^*u_1).
\end{equation}
Since only the imaginary part of $C(\omega)$ is odd with respect to frequency, only this piece will contribute to $A_1,A_2$ and those parameters
will therefore be real quantities.  The final step is to compute the temperature dependence  of $C'(\omega=0)$ in the spin-incoherent regime.

\begin{figure}[h]
\includegraphics[width=.75\linewidth,clip=]{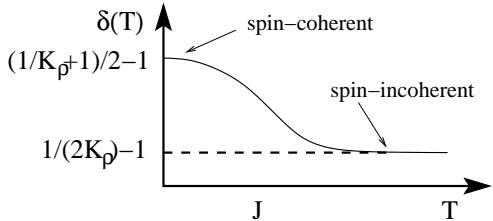}
 \caption{\label{fig:delta} Temperature dependence of the exponent $\delta$ appearing in the parameters \eqref{eq:A_temp} that describe the pumped
 spin current \eqref{spnexpr} due to time-dependent motion of a magnetization vector near an interacting acting one-dimensional system at finite
 temperature.  For temperatures $T$ less than spin exchange $J=E_\sigma$, the characteristic spin pumping temperature dependence in the spin-incoherent
 regime crosses over to the Luttinger liquid form computed in Ref.[\onlinecite{Bena:prb04}]. Note that in the SILL, $\delta(T)<0$ for $K_\rho>1/2$ and
 the qualitative temperature dependence of the pumped spin current is remarkably different from a LL. }
 \end{figure}

Since we are working within linear response, the commutator in $C(t) =
-i\Theta(t)\langle[F^\dagger_s(t)\psi_s(t),\psi^\dagger_s(0)F_s(0)]\rangle$ is evaluated in the state where there is no tunneling between the
ferromagnet and the SILL, and a hard-wall at $x=0$.  By writing $C(t)=-i\Theta(t)[\langle F_s^\dagger(t) F_s(0)\rangle \langle
\psi_s(t)\psi_s^\dagger(0)\rangle - \langle F_s(0) F_s^\dagger(t)\rangle   \langle \psi_s^\dagger(0)\psi_s(t)\rangle]$, and noting that the long
time behavior of both terms has the same functional dependence, we can easily extract the long time behavior\cite{Fiete:rmp07} of $C(t)$ using
\eqref{eq:S_FM} and \eqref{eq:S_SILL}. (A more careful calculation that yields the same result is given in Appendix~\ref{app:spin_current}.) At
finite temperatures (small with respect to the Fermi energy of the ferromagnet and charge energy of the SILL, but large compared to $\hbar
v_\rho/L$), we have $\langle F_s^\dagger(t) F_s(0)\rangle \sim \frac{(\pi k_B T/\hbar)}{\sinh(\pi k_B T t/\hbar)}$ and $\langle
\psi_s(t)\psi_s^\dagger(0)\rangle \sim  \left[\frac{(\pi k_B T/\hbar)}{\sinh(\pi k_B T t/\hbar)}\right]^{1/(2K_\rho)}$. The crucial difference
with the Luttinger liquid is that the exponent for the correlations $\langle \psi_s(t)\psi_s^\dagger(0)\rangle$ have changed: in the case of a
LL $1/(2K_\rho)$ is replaced by $(1/K_\rho+1)/2$.  The remainder of the computations carry through exactly as they would for a LL and we find
for $\hbar \omega \ll k_B T$
\begin{equation}
\label{eq:C_omega_sketch}
{\rm Im}[C(\omega)] \propto \hbar \omega \left(k_B T\right)^{\delta^{SILL}},
\end{equation}
where $\delta^{SILL}=\frac{1}{2K_\rho}-1$.   This implies $C'(0) \propto T^{\delta(T)}$ quite generally so that
\begin{equation}
\label{eq:A_temp}
A_1\propto T^{\delta(T)} \frac{|u_2|^2}{2},\;\;  A_2 \propto T^{\delta(T)} {\rm Im}(u_2^*u_1),
\end{equation}
where $\delta(T)$ interpolates between the LL and SILL regimes as the temperature is swept through the spin energy $E_\sigma \approx J$. The
temperature dependence of the exponent $\delta(T)$ is shown schematically in Fig.~\ref{fig:delta}.

It is interesting to note that $\delta>0$ for $0<K_\rho<1$ in the Luttinger liquid regime implying that less spin current is pumped as the
temperature decreases.  On the other hand, so long as the system remains in the spin-incoherent regime the opposite behavior may be obtained if
$K_\rho>1/2$: since $\delta^{SILL}<0$ {\em more} spin current is pumped as the temperature is lowered (for $T \gtrsim J$).  This is related to
the diverging density of states at the boundary\cite{Fiete:prb05,Fiete:rmp07,Kindermann_crossover:prb06} when $K_\rho>1/2$.  In gated
cleaved-edge overgrowth quantum wires it appears possible to lower $K_\rho$ down to values of order $1/3$ and so it may be possible in the
spin-incoherent regime to tune between the positive and negative exponent regimes.\cite{Auslaender:sci05}  This qualitative difference should be
easily seen in experiment.

\subsection{Renormalization of Magnetization Dynamics}

Having computed the spin current pumped into the SILL by a time-dependent magnetization vector, it is important to ask how the spin accumulation
in the SILL in turn affects the magnetization motion. We address this question by computing the renormalization of the Gilbert damping constant
due to the spin flow into the SILL.  We again closely follow the notation of Bena and Balents\cite{Bena:prb04} to clearly establish a connection
to the LL case.  The Landau-Lifshitz-Gilbert equation for magnetization $\vec m$ precessing around effective magnetic field $\vec H$ is
\begin{equation}
\frac{d \vec m}{dt}=-\gamma \vec m \times \vec H+ \alpha_0 \vec m \times \frac{d \vec m}{dt}-\frac{\gamma}{M_s} \vec I,
\label{eq:m_dyn}
\end{equation}
where the spin current $\vec I=\vec I_0-\vec I_b$ flows ($\vec I_0$ is given by \eqref{spnexpr} for $\vec I_b \equiv 0$ and is non-zero only for
time dependent $\vec m$) into the SILL and $M_s$ is the saturation magnetization of the ferromagnet.  Here $\gamma$ is the gyromagnetic ratio
which is typically equal to its free electron value,\cite{Tserkovnyak:rmp05} $\gamma=2\mu_B/\hbar$, in transition metal ferromagnets, and
$\alpha_0$ is the dimensionless Gilbert damping parameter in the absence of the spin current.  Its value is typically of order $10^{-2}$. The
spin backflow due to spin accumulation in the interacting 1-d system can be described by boundary conditions on left and right moving spin
currents and is expressed as\cite{Balents:prb01,Balents:prl00}
\begin{equation}
\label{eq:I_b}
\vec I_b=\frac{\vec \mu_s}{\mu_s} I_{\delta(T)}(\mu_s,T)-\frac{K_{\rm exch}}{4\pi} \vec \mu_s \times \vec m,
\end{equation}
where $\vec \mu_s$ is the spin chemical potential in the wire related to the magnetization by $\vec \mu_s(x)=\vec M(x)/\chi(T)$, with $\chi(T)$
the (generally temperature dependent) spin susceptibility, and $K_{\rm exch}$ describes the effective exchange coupling between electrons in the
SILL and the FM.\cite{RG_remark}  The current  $I_{\delta(T)}(\mu_s,T)$ arises from the electron tunneling contribution to the spin
current.\cite{Si:prl98}  Note that the back-scattered spin current \eqref{eq:I_b} contains two terms: (i) a term arising from electron transfer
from the FM to the interacting 1-d system and (ii) a second term arising from exchange between the local spin density and the precessing
magnetization vector $\vec m$. Since $\vec I=\vec I_0-\vec I_b$ this implies the spin current itself has a contribution due purely to exchange
effects, even in the absence of electron transfer.\cite{Balents:prl00}

Our goal in this section is to  express \eqref{eq:m_dyn} as
\begin{equation}
\frac{d \vec m}{dt}=-\gamma' \vec m \times \vec H+ \alpha \vec m \times \frac{d \vec m}{dt},
\end{equation}
and determine the renormalized parameters $\gamma'$ and $\alpha$. From \eqref{eq:m_dyn} it is evident that we must find $\vec I =\vec I_0-\vec
I_b$.  We have already computed $\vec I_0$, Eq.\eqref{spnexpr}, in the previous subsection. Now we must determine $\vec I_b$, \eqref{eq:I_b},
which depends on $\vec \mu_s$ and $I_{\delta(T)}(\mu_s,T)$.  We start with $\vec \mu_s$ which is a function of the spin dynamics in the interacting
1-d system.

It has earlier been shown that SU(2) invariant electron backscattering leads to diffusive spin behavior in the weakly interacting
regime.\cite{Bena:prb02,Balents:prb01}  In the strongly interacting regime where the spin sector may effectively be described by a Heisenberg
spin chain coupled to phonon distortions\cite{Fiete:rmp07,Fiete:prb06,Matveev:prb04,Matveev:prb07} the spin dynamics has also been shown to be
diffusive.\cite{Narozhny:prb96}   In both limits, the mean free path $l\sim v_\sigma/T$.  Following the
arguments\cite{Fiete_2:prb05,Iucci:prb07} that the spin-incoherent regime can be approached from ``below" ($T < J$), we expect the diffusion
length $l$ to saturate at the interparticle spacing, $a$, for $T \gtrsim J$.  Since the diffusion constant $D_s\sim l v_\sigma$, this implies
the diffusion constant in the SILL becomes independent of temperature and takes the value $D_s\approx a v_\sigma$.  Note that in the limit
$v_\sigma \to 0$,  there is no spin diffusion, as the spin excitations cannot propagate in the system. Spin will simply pile up at the boundary
of the FM without moving further into the interacting 1-d system.  Unlike our earlier results for the temperature dependence of pumped spin into
the SILL, Eq.\eqref{eq:A_temp} with $\delta^{SILL}=1/(2K_\rho)-1$, which were valid for vanishing spin velocity, here our results are
qualitatively dependent on keeping $v_\sigma$ finite, although still small enough to be in the spin-incoherent regime.

We also assume the finite length SILL is characterized by a spin-flip time $\tau_{sf}$ (due to impurities, spin-orbit effects, etc). The
diffusion equation for spin in the SILL is then
\begin{equation}
i \omega \vec \mu_s(x)=D_s \partial^2_x \vec \mu_s(x) -\tau_{sf}^{-1}\vec \mu_s(x),
\end{equation}
with the boundary conditions $\partial_x \vec \mu_s=-\left(\frac{1}{D_s \chi(T)}\right) \vec I$ at $x=0$ and vanishing spin current $\partial_x
\vec \mu_s(x)=0$ at $x=L$.  See  Fig.~\ref{fig:FM} for the set-up.  The solution to this equation is simple to obtain and is given by
\begin{equation}
\vec \mu_s(x) = \frac{\cosh[\kappa (L-x)]}{\kappa \sinh[\kappa L]}\left(\frac{1}{D_s \chi(T)}\right) \vec I(x=0),
\end{equation}
where $\kappa = \sqrt{\frac{i\omega-\tau_{sf}^{-1}}{D_s}}$.  If the precession frequency $\omega$ is small compared to the inverse spin
relaxation time, then to a good approximation $\kappa \approx 1/\sqrt{D_s \tau_{sf}}$. At the boundary $x=0$ we have $\vec \mu_s=\xi  \vec I$
where $\xi=\coth[\kappa L] \frac{1}{\kappa D_s \chi(T)}$.  Note that  $\partial_x\vec \mu_s(x) \sim e^{-\kappa x}\vec I(x=0) \approx
e^{-x/\sqrt{D_s \tau_{sf}}}\vec I(x=0)$  so the spin current decays exponentially with distance into the SILL on a length scale set by the
product of the diffusion constant and the spin-relaxation time.  For a long spin-relaxation time, this length scale can be large compared to the
inter-particle spacing.

We have now determined all parts of the spin back flow \eqref{eq:I_b}, except the tunneling current $I_{\delta(T)}(\mu_s,T)$, which we now do.
The tunneling current is proportional to the imaginary part of the Fourier transformed correlation function\cite{Balents:prl00,Balents:prb01}
$C(\omega)$ defined below Eq.~\eqref{spncurntfnl},
\begin{eqnarray}
I_{\delta(T)}(\mu_s,T) \propto |u_1|^2 {\rm Im}[C(\mu_s/2)]\nonumber \\
 \propto |u_1|^2(k_B T)^{\delta(T)+1} \sinh\left(\frac{\mu_s}{4 k_B T}\right) \nonumber \\
\times \left|\Gamma\left(1+\frac{\delta(T)}{2}+i\frac{\mu_s}{4\pi k_B T}\right)\right|^2 \nonumber \\
\propto \mu_s |u_1|^2(k_B T)^{\delta(T)}\left|\Gamma\left(1+\frac{\delta(T)}{2}\right)\right|^2,
\end{eqnarray}
where we have used the results of Appendix \ref{app:spin_current} and taken the limit $\mu_s \ll k_B T$. Evidently, the main effect of the
spin-incoherent physics is to change the exponent $\delta(T)$ to the spin-incoherent value, $\delta^{SILL}=1/(2K_\rho)-1$, so that we have
\begin{equation}
I_{\delta(T)}(\mu_s,T) \propto \mu_s |u_1|^2 (k_B T)^{1/(2K_\rho)-1}
\end{equation}
in the spin-incoherent regime.  Let us define ${\cal T}\equiv I_{\delta(T)}(\mu_s,T)/\mu_s \propto (k_B T)^{\delta^{SILL}}$ which is temperature
dependent and $\mu_s$ independent.  At this point the determination of $\gamma'$ and $\alpha$ is identical to the LL case.\cite{Bena:prb04}  We
find
\begin{equation}
\gamma'=\frac{\gamma}{1+\gamma(B_1A_2-B_2A_1)/M_s},
\end{equation}
and
\begin{equation}
\alpha=\frac{\alpha_0+\gamma(B_1A_1+B_2A_2)/M_s}{1+\gamma(B_1A_2-B_2A_1)/M_s},
\end{equation}
where $B_1=(1+\xi {\cal T})/[(1+\xi {\cal T})^2+(\xi K_{\rm exch})^2/(16 \pi^2)]$, $B_2=(\xi K_{\rm exch})/[(1+\xi {\cal T})^2+(\xi K_{\rm
exch})^2/(16 \pi^2)]$, and $A_1,A_2$ are given in Eqs.(\ref{eq:A_gen},\ref{eq:A_temp}).  The temperature dependence in $B_1,B_2$ is entirely
contained in ${\cal T}$ and $\xi$.  See Table~\ref{tb:LL_SILL}  below for a comparison of the LL and SILL regimes.

\begin{table} [h]
\caption{Temperature dependence of key quantities in LL-FM and SILL-FM hybrid systems.  Here $c'$ is a constant that depends on $L,\tau_{sf}$,
and $D_s$.} \label{tb:LL_SILL}
\begin{tabular}{| c | c | c | c | c | c  |}
\hline
& $\delta$ & $D_s$ & $\chi$ & $\xi$ & ${\cal T},A_1, A_2$ \\
\hline
SILL & ${1\over 2 K\rho}-1$ & const & $ {1 \over T}$ & $ T$ & $ T^{
\delta^{SILL}}$\\
\hline
LL & $\frac{1}{2} \left({1\over  K\rho}+1\right)-1$ & $ {1 \over T}$& const  & $ \coth(c' \sqrt{T}) \sqrt{T}$ & $T^{\delta^{LL}}$\\
\hline
\end{tabular}
\end{table}

As with the case of a LL, for a SILL we expect there to be little renormalization of $\gamma'$ relative to $\gamma$, but the smallness of the
Gilbert damping $\alpha_0$ means this may obtain a significant temperature dependent correction depending on $\delta^{SILL}$.

\subsection{Other FM hybrid structures involving a SILL}

One could easily imagine other scenarios such as a FM-SILL-FM junction, a FM-SILL-M junction, or even a FM-SILL-SC junction.  However, because
the spin transport is generally diffusive, for junctions whose length $L$ is long compared to $\sqrt{D_s \tau_{sf}}$, the two contacts to the
SILL will more or less behave independently from the point of view of spin transport.   To the extent that the two leads are coupled, it is
evident from the discussion of the previous section that most of the physics (and corresponding general formulas) present for a LL system  also
apply to the SILL system only with some modifications in the temperature dependence of parameters appearing in the theory.  While this may seem
like a somewhat trivial result, it is not.  The temperature dependence serves as a means to determine whether spin-incoherent physics is present
in the system.  In particular, we note quite generally that if $1/2<K_\rho <1$ the temperature dependence of many quantities (e.g., ${\cal
T},A_1,A_2$ and those derived from them) change {\em qualitatively} relative to the expectations for a LL.  Such qualitative differences should
be observable in experiment.

It is worthwhile to step back and emphasize some of the essential differences between the SILL and LL cases. First, we note that the different
temperature dependence  can be traced to three quantities:  (1) The correlation function $C(t)$ defined below Eq.~\eqref{spncurntfnl}. (2) The
temperature dependence of the spin susceptibility $\chi$.  (3) The temperature dependence of the spin diffusion constant $D_s$.  So long as the
SILL is tunnel contacted to the FM, the correlation function $C(t)$ will always appear at lowest order in perturbation theory and carry along
with it the characteristic temperature dependence of the SILL.  This quantity appears in both the description of the pumped spin current and the
magnetization dynamics.  On the other hand, when we are specifically interested in how the spin propagates in the SILL [as we saw for the
backscattered spin current \eqref{eq:I_b}] the diffusion constant $D_s$ enters, and also the spin susceptibility via the Einstein relation for
the spin conductivity, $\sigma_s=\chi D_s$.  We remark that the spin transport is generically diffusive in the spin-incoherent regime, while it
may be either ballistic or diffusive in the LL regime.\cite{Bena:prb02,Balents:prb01}

Having now flushed out what we feel are the central considerations and results for FM-SILL hybrid structures, we now turn our attention to
SC-SILL hybrid structures.

\section{SC-SILL Hybrid Structures}
\label{sec:SC-SILL}

Throughout this work, we assume that all energy scales are small compared to the superconducting gap, $\Delta$, unless explicitly stated otherwise, such as in the limit of a short junction (defined below).  We are primarily  interested in the case where the superconductor is contacted via a tunnel junction with a spin-incoherent Luttinger liquid.

\subsection{SC-SILL Junctions}

Let us begin our discussion of SC-SILL hybrid systems by considering the simplest case: where the FM in Fig.~\ref{fig:FM} is replaced by a SC.
Earlier we studied the properties of such a junction in the Andreev limit.\cite{Tilahun:prb08}  We found a number of remarkable properties,
including a completely universal (independent of the charge and spin Hamiltonians) tunneling density of states in the spin-incoherent
regime.\cite{Tilahun:prb08}  In the opposite limit of a tunnel contact between the SC and SILL, no proximity effects of the SC are felt other
than those perturbative in the tunneling.\cite{Fazio:sm99} (In contrast to the Andreev limit where the behavior of the pair correlations and
tunneling density of states follows directly from the form of the field expansions imposed by the boundary conditions.\cite{Tilahun:prb08}) As
the proximity effects are already weak in the Andreev limit (they decay exponentially fast with distance, unlike the power law decay in the
LL\cite{Takane:jpsj96}) we do not pursue the very weak proximity effects in the spin-incoherent regime in the limit of a weak tunneling SC-SILL
junction.

On the other hand, it is worthwhile to briefly discuss the behavior expected in opposite limit of nearly perfect Andreev reflection with weak
(perturbative) electron backscattering at the SC-SILL interface.  In the LL context, this was addressed by Vishveshwara {\it et
al.}\cite{Vishveshwara:prb02} in a SC-LL-M system who found dips in the conductance of the junction as a function of gate voltage. These dips
correspond to multiple electron reflections off the SC interface and occur at voltages corresponding to integer values, $eV=n \hbar \pi/\tau$, of the inverse electron traversal
time, $\tau=2L/v$,  where $n$ is an integer counting the number of traversals in the LL.\cite{Vishveshwara:prb02} For a
SC-SILL-M system we would expect similar qualitative effects, although with an overall suppression in the
conductance.\cite{Matveev:prl04,Matveev:prb04,Tilahun:prb08} Finally, we briefly remark that the conductance per channel of an adiabatic SC-SILL
junction should drop to 1/2 the value expected for a SC-LL junction:\cite{Takane:jpsj96,Lee:prl03} $G_{SC-SILL}=\frac{1}{2} (2K_\rho)
\frac{2e^2}{h}$.  Recall that the conductance of the adiabatic SC-LL interface $G_{SC-LL}=(2K_\rho) \frac{2e^2}{h}$ generalizes the result for
the adiabatic SC-M interface\cite{Blonder:prb82} $G_{SC-M}=2 \frac{2e^2}{h}$ realized for $K_\rho=1$.

\subsection{SC-SILL-SC Tunnel Junctions}
\label{sec:SC-SILL-SC}

\begin{figure}[h]
\includegraphics[width=.75\linewidth,clip=]{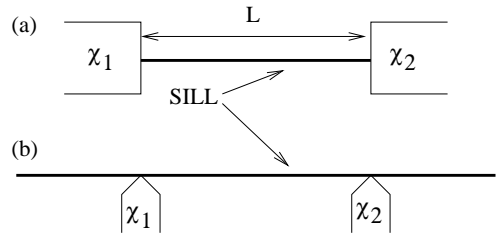}
\caption{\label{fig:SC_tunnel} Schematic of  two ways  a superconductor can contact a spin-incoherent Luttinger liquid of length $L$. The phase
of the superconducting order parameter at the left contact is $\chi_1$ and at the right contact $\chi_2$.  We assume there is a tunnel
barrier between the SC and SILL.  In (a) the SILL is ``end contacted" and the zero modes (finite length) of the SILL generally come into play,
while in (b) the SILL is ``bulk contacted" and the zero modes play no role since the system length is effectively infinite.}
 \end{figure}

We now turn our attention to the final topic of this work, SC-SILL-SC junctions of the types shown in Fig.~\ref{fig:SC_tunnel}. Our primary
focus will be on the behavior of the ac and dc Josephson effects in such a hybrid structure. In the context of SC-LL-SC structures there have
already been a number of theoretical
works\cite{Maslov:prb96,Fazio:prb96,Takane:jpsj97,Fazio:prl95,Affleck:prb00,Caux:prl02,Feiguin:prb07,Yokoshi:prb07,Yokoshi:prb05,Krive:prb05}
and some experimental results.\cite{Doh:sci05,Lindelof:march07,Xiang:natn06,Kasumov:prb03,Morpurgo:sci99}

Earlier we studied the dc Josephson effect in a junction like that shown in Fig.~\ref{fig:SC_tunnel}(a) only with Andreev boundary conditions at
the interfaces of the SILL with the SCs.\cite{Tilahun:prb08}  We found that in spite of the fact that the pair correlations decay exponentially
with distance from the boundary of the SC, the Josephson critical current scaled inversely with the length of the SILL and was reduced by a
factor of 2 relative to either the SC-LL-SC or SC-M-SC case (which have identical critical currents that also scale inversely with the
length).\cite{Maslov:prb96,Affleck:prb00}  We remarked that this ``robustness" is essentially related to the fact that the superconductor phase
difference only couples to the charge degrees of freedom in the SILL, which remain completely coherent.    This effect can also be taken to
support the notion\cite{Kivelson:prb90} that in some sense superconductors are spin-charge separated.\cite{LeHur_comment}  Yet another perspective
on this result can be obtained by noting that the Andreev boundary conditions on a system of finite length can be mapped onto a system with {\em
periodic} boundary conditions of twice the length.\cite{Maslov:prb96}  In this way, the Josephson response becomes equivalent to the problem of
persistent currents for {\em spinless} electrons in a ring threaded by magnetic flux, with the superconductor phase difference being simply
related to the flux.  Therefore, the dc Josephson current for Andreev boundary conditions is essentially given by the physics of {\em spinless}
electrons which is why spin-incoherent physics does not manifest in a dramatic way.  The factor of 2 reduction in the critical current and
Josephson period are a direct consequence of the mapping of the SILL onto spinless electrons.

We now turn our attention to the case of open boundary conditions at the interfaces between the SCs and SILL. In this case phase information
must be carried between the two superconductors by Cooper pair tunneling between them. Because the {\em pairs} must hop from one SC onto the SILL,
then onto the other SC, the Josephson current will occur at 4th order in
this tunneling process.  Moreover, because the Cooper pair tunneling process creates a pair of electrons locally in the SILL near the barrier,
many Fourier modes of the electron operator will come into play.  This is in contrast to the case of Andreev boundary conditions where only the
zero mode enters in the evaluation of   Josephson current.\cite{Tilahun:prb08,Maslov:prb96}  This crucial difference will turn out to
dramatically suppress the Josephson response  relative to what one would have for a LL when the SILL is tunnel coupled to SCs. In short, we find the critical current decays as an exponential of the junction length (compared to a power of the inverse junction length for a LL at zero temperature) with the characteristic decay length the inter-particle spacing.

\subsubsection{DC Josephson Effect}
\label{sec:DC-Josephson}

The SC-SILL-SC tunnel junctions shown in Fig.~\ref{fig:SC_tunnel} are modeled by the following Hamiltonian, $H = H_{S1} + H_{S2} + H_{SILL} +
H_T$ where $H_{S1/2}$ are $s$-wave BCS Hamiltonians for the superconductors and $H_{SILL}$ is the sum of Eqs. \eqref{eq:H_holon} and
\eqref{eq:S_chain}. The tunneling Hamiltonian $H_T$ is given by
\begin{eqnarray}
H_T &=& \sum_s T_1\psi_s^{S1}(x=0)\psi_s(x=0)\nonumber\\&+&T_2\psi_s^{S2}(x=L)\psi_s(x=L)+h.c. \label{eq:tunHamSC}
\end{eqnarray}
We assume the SCs support a phase difference of $\chi = \chi_2-\chi_1$ and have the same superconducting gap $\Delta$.

The Josephson current is obtained from the relation 
\begin{equation}
\mathcal{J} = -2ek_B T\frac{\partial \ln Z}{\partial \chi},
\end{equation}
 where $-e$ is the electronic
charge and $Z$ is the partition function. We employ imaginary time perturbation theory, where terms dependent on $\chi$ appear in fourth
order,\cite{Fazio:prb96}
\begin{eqnarray}
\ln Z = \int d\tau_1d\tau_2d\tau_3d\tau_4
T^2_1(T^*_2)^2F^\dagger_{S1}(\tau_1-\tau_2)\times\nonumber\\\Pi(0,L;\tau_1,\tau_2,\tau_3,\tau_4)F_{S2}(\tau_3-\tau_4)+\nonumber\\
h.c.+\text{similar terms},\nonumber\\\label{eq:frengy}
\end{eqnarray}
where $\Pi(0,L;\tau_1,\tau_2,\tau_3,\tau_4)$ is a two-particle (Cooperon) propagator in the one-dimensional interacting system. The ``similar
terms" account for time ordered permutations and spin projections (there are 23 of them, $4!$ all together). The propagation of Cooper pairs in the
superconductors is described by the time-ordered anomalous Green function $F_S$,
\begin{eqnarray}
F_{S_i}(\tau-\tau ') &\equiv& \langle T_\tau\psi_{\downarrow}^{S_i}(\tau)\psi_{\uparrow}^{S_i}(\tau')\rangle\nonumber\\&=& \frac{\pi
N_i(0)}{\beta}\sum_ne^{-i\omega_n(\tau-\tau ')}\frac{\Delta e^{i\chi_{S_i}}}{\sqrt{\omega^2_n+\Delta^2}},\nonumber \\ 
\label{eq:anmlgrnfn}
\end{eqnarray}
where the expectation value is taken with respect to $H_{Si}$ and the electron operators in \eqref{eq:anmlgrnfn} are evaluted at $x=0$ for $S_1$ and $x=L$ for $S_2$. 

Of the 24 terms (plus their Hermitian conjugates) that appear in the partition function, Eq.~\eqref{eq:frengy}, physical considerations aid us in choosing the most relevant ones depending on the length of the SC separation distance $L$. There are two important length scales in the junctions we study that give rise to corresponding time scales: The superconducting coherence length $\xi$ and the junction length $L$.\cite{coherence_length_remark}  We define ``long"  to mean $L \gg \xi$ and ``short" to mean $L \ll \xi$. We now turn to a discussion of the general expression for $\Pi(0,L;\tau_1,\tau_2,\tau_3,\tau_4)$ in the limit of long and short junctions.

For long junctions the tunneling into the SILL is ``fast" and the propagation of the Cooper pairs is ``slow".  The relevant two-body propagator is 
\begin{eqnarray}
\Pi^{\rm long}(0,L;\tau_1,\tau_2,\tau_3,\tau_4) = \nonumber\\\left \langle
\psi_s(0,\tau_1)\psi_{-s}(0,\tau_2)\psi^{\dagger}_{-s}(L,\tau_3)\psi^{\dagger}_{s}(L,\tau_4)\right\rangle,
\nonumber\\
\label{eq:SILL-propgt-Long}
\end{eqnarray}
where $\psi_s$ is the bosonized electron annihilation operator given by Eq. \eqref{eq:psi_final_bos} and the averaging is taken with respect to
$H_{SILL}$. We have $|\tau_1-\tau_2| \propto \xi$, $|\tau_3-\tau_4| \propto \xi$, and $|\tau_1 -\tau_4|\approx |\tau_2 - \tau_3| \propto L$.

In the opposite limit of a short junction $L \ll \xi$, the propagation through the SILL is ``fast" compared to the ``slow" tunneling of the Cooper pairs, so the two-particle propagator approximately separates into a product of two single-particle Green's functions\cite{Fazio:prb96}
\begin{eqnarray}
\Pi^{\text{short}}(0,L;\tau_1,\tau_2,\tau_3,\tau_4) 
\nonumber \\
\approx \left\langle\psi_{\vphantom{-}s}(0,\tau_1)
\psi^{\dagger}_{\vphantom{-}s}(L,\tau_2)\right\rangle \left\langle\psi_{-s}(0,\tau_3)\psi^{\dagger}_{-s}(L,\tau_4)\right\rangle\nonumber \\ \approx G_{\vphantom{-}s}(L,\tau_1-\tau_2)G_{-s}(L,\tau_3-\tau_4), 
\label{eq:SILL-propgt-Short}
\end{eqnarray}
where we have $|\tau_1-\tau_2| \propto \xi$, $|\tau_3-\tau_4| \propto \xi$, and $|\tau_1 -\tau_3|\approx |\tau_2 - \tau_4| \propto L$ (note the re-ordering of electron operators relative to \eqref{eq:SILL-propgt-Long}).

{\em End contacted SILL when $L\gg \xi$}:
We first consider the case of an ``end-contacted" SILL, shown in Fig.\ref{fig:SC_tunnel}a. For this geometry the expansions \eqref{eq:open_expansions} of the holon field operators are the appropriate ones. Our assumption that all energy scales are small compared to $\Delta$ (specifically
$k_BT \ll \Delta$, and $\hbar v_\rho/L \ll \Delta$) enables us to approximate the anomalous Green's function  \eqref{eq:anmlgrnfn} with a delta function in time, $F_{S_i}(\tau-\tau') \approx \pi N_i(0)
e^{i\chi_i}\delta(\tau-\tau')$ where $N_i(0)$ is the density of states of the SC $S_i$ at the Fermi energy in its normal state. This approximation forces the times the electrons tunnel into or out of the SILL to
coincide, $\tau_1=\tau_2$ and $\tau_3=\tau_4$. Without loss of generality, we let $\tau_3=0$ and define $\tau\equiv\tau_1$. Using the expression
for the electron operator, Eq.\eqref{eq:psi_final_bos}, the propagator \eqref{eq:SILL-propgt-Long} reads
\begin{widetext}
\begin{eqnarray}
\Pi^{\rm{ long}}(0,L;\tau,0) &=& \left(\frac{1}{2\pi \alpha_c}\right)^2 \sum_{l_1,l_2,l_3,l_4}\int_{-\infty}^\infty
\frac{dq_1}{2\pi}\int_{-\infty}^\infty \frac{dq_2}{2\pi}\int_{-\infty}^\infty \frac{dq_3}{2\pi}\int_{-\infty}^\infty \frac{dq_4}{2\pi}
e^{-i\left(q_1l_1+q_2l_2-q_3l_3-q_4l_4\right)}\nonumber\\&\vphantom{\times}&\bigg\langle
e^{-i\phi(0,\tau)}Z^{\vphantom{\dagger}}_{l_1,\vphantom{-}s}e^{i\left(1+\frac{q_1}{\pi}\right)\left(k_F^h\cdot
0+\theta(0,\tau)\right)}e^{-i\phi(0,\tau)}Z^{\vphantom{\dagger}}_{l_2,-s}e^{i\left(1+\frac{q_2}{\pi}\right)\left(k_F^h\cdot
0+\theta(0,\tau)\right)}
\nonumber\\&\times&e^{-i\left(1+\frac{q_3}{\pi}\right)\left(k_F^hL+\theta(L,0)\right)}
Z^{\dagger}_{l_3,-s}e^{i\phi(L,0)}
e^{-i\left(1+\frac{q_4}{\pi}\right)\left(k_F^hL+\theta(L,0)\right)}Z^{\dagger}_{l_4,\vphantom{-}s}e^{i\phi(L,0)}\bigg\rangle.
\label{eq:PiwithZ}
\end{eqnarray}
\end{widetext}
The calculation of the Cooper pair propagator is carried out in the same manner as that of the Green's function presented in Appendix \ref{app:spin_current}. We first evaluate the expectation value of the spin chain site creation/annihilation operators, 
\begin{equation}
\Xi(0,L) \equiv \left\langle Z^{\vphantom{\dagger}}_{l_1,\vphantom{-}s}Z^{\vphantom{\dagger}}_{l_2,-s}
Z^{\dagger}_{l_3,-s}Z^{\dagger}_{l_4,\vphantom{-}s}\right\rangle,
\label{eq:zhi}
\end{equation}
in the spin-incoherent regime. To this end it is convenient to fully exploit the symmetries and boundary conditions of the problem. We first note that the open boundary conditions at $x=0,L$ force $\theta=$const at both these points. As before, we take this constant to be zero so that all $\theta$ fields effectively drop out of Eq.\eqref{eq:PiwithZ}. The momentum integrals in \eqref{eq:PiwithZ} can then be done trivially to give delta functions on the sites $l_i$ which sets $l_1=l_2=0$ and $l_3=l_4=k_F^h L/\pi$.  Therefore, we have   
\begin{equation}
\Xi(0,L) =\left\langle Z^{\vphantom{\dagger}}_{0,\vphantom{-}s}Z^{\vphantom{\dagger}}_{0,-s}
Z^{\dagger}_{k_F^hL/\pi,-s}Z^{\dagger}_{k_F^hL/\pi,\vphantom{-}s}\right\rangle,
\label{eq:zhi_simplify}
\end{equation}
which in the spin-incoherent regime can be evaluated quite simply by appealing to the physics of the spins. Here $\Xi(0,L)$  measures the amplitude for two spins of opposite orientation introduced at $x=L$ to arrive at $x=0$. Deep in the SILL regime where all the spins are randomized and spin exchange is highly suppressed, this is just the probability of
finding neighboring sites of the spin chain housing oppositely aligned spins,
\begin{eqnarray}
\Xi(0,L) = \left(\frac{1}{2}\right)^{k_F^hL/\pi}=e^{-\frac{k_F^hL}{\pi} \ln2},
\label{eq:ZhiVal}
\end{eqnarray}
from which it follows that
\begin{equation}
\Pi^{\rm{ long}}_{\rm end}(0,L;\tau,0)=e^{-\frac{k_F^hL}{\pi} \ln2} \langle e^{-i 2\phi(0,\tau)}e^{i2\phi(L,0)}\rangle.
\end{equation} 
The remaining correlations over the charge degrees of freedom can be computed using the expansions \eqref{eq:open_expansions} and the identity $e^{A+B}=e^A e^B e^{-[A,B]/2}$ valid when $[A,B]$ commutes with both $A$ and $B$.  The final result is
\begin{equation}
\Pi^{\rm{ long}}_{\rm end}(0,L;\tau,0)=e^{-\frac{k_F^hL}{\pi} \ln2}  \left(\frac{\eta}{1+e^{-\eta}e^{-v_\rho \pi \tau/L}}\right)^{\frac{1}{K_\rho}}
\end{equation} 
where $\eta \ll 1$ is a short distance cut-off that ensures ultra violet convergence of the integrals over the charge fluctuations in the wire.  Up to unimportant phase factors and other overall multiplicative constants, the Josephson current in the wire is then
\begin{eqnarray}
{\cal J}^{\rm long}_{\rm end}(\chi) \propto G_1 G_2 e^{-\frac{k_F^hL}{\pi} \ln2} \left(\frac{L}{\alpha_c}\right)\sin(\chi)\nonumber \\
\times \int_0^\pi dx  \left(\frac{\eta}{1+e^{-\eta}e^{-x}}\right)^{\frac{1}{K_\rho}},
\label{eq:endJosCurntLong}
\end{eqnarray}
where the integral has been cut off at times $\tau =L/v_\rho$, the time it takes for charge to propagate between the two ends of the SILL.
Compared to the LL case, the most important difference is that the critical current scales as an exponential of the length, ${\cal J}^{\rm long}_{\rm end}|_{\rm critical} \propto e^{-\frac{k_F^hL}{\pi} \ln2} \left(\frac{L}{\alpha_c}\right)$ rather than a power law $\left(\frac{\alpha_c}{L}\right)^{\frac{2}{K_\rho}-1}$ in the LL regime.\cite{Takane:jpsj97}  A measurement of the length dependence of the critical current then serves as clear signature of spin-incoherent physics (or lack thereof), provided one has some knowledge of the density to infer $k_F^h=2k_F=\pi/a$.  We also note that the superconducting phase difference $\chi$ appears as an argument to the $\sin$ function, rather than the saw-tooth form we found for Andreev contacts.\cite{Tilahun:prb08}  The $\sin$ form follows from the assumption of tunneling contacts.\cite{Caux:prl02,Feiguin:prb07,Affleck:prb00}

{\em Bulk contacted SILL when $L \gg \xi$}:
We now consider the case of a ``bulk-contacted" SILL, shown in Fig.\ref{fig:SC_tunnel}b. For this case, we assume that the perturbations induced by the contacts are irrelevant, which is so for $K_\rho >1/2$.\cite{Fiete_2:prb05}  Otherwise, if the perturbations induced by the contacts are relevant this case reduces to an ``end-contacted" SILL discussed above.  

Starting again with \eqref{eq:PiwithZ} we note that translational symmetry implies that the spin correlations satisfy
\begin{eqnarray}
\Xi(0,L) &\equiv &\left\langle Z^{\vphantom{\dagger}}_{l_1,\vphantom{-}s}Z^{\vphantom{\dagger}}_{l_2,-s}
Z^{\dagger}_{l_3,-s}Z^{\dagger}_{l_4,\vphantom{-}s}\right\rangle \nonumber \\
&=&\left\langle Z^{\vphantom{\dagger}}_{l_1-l_4,\vphantom{-}s}Z^{\vphantom{\dagger}}_{l_2-l_3,-s}
Z^{\dagger}_{0,-s}Z^{\dagger}_{0,\vphantom{-}s}\right\rangle,
\label{eq:zhi_trans}
\end{eqnarray}
which allows us to shift the sums: $\tilde l_1=l_1-l_4, \tilde l_2=l_2-l_3$ in \eqref{eq:PiwithZ}.  The summation over $l_3$ leads to a delta function setting $q_2=q_3$ and the summation over $l_4$ leads to a delta function setting $q_1=q_4$.  Making the change of variables $q=(q_1+q_2)/2$ and $\tilde q=q_1-q_2$ one can do the integration over $\tilde q$ which sets $\tilde l_1=\tilde l_2$.  The result is
\begin{eqnarray}
\Pi^{\rm{ long}}_{\rm bulk}(0,L;\tau,0) =\left(\frac{1}{2\pi \alpha_c}\right)^2\sum_l \int_{-\infty}^\infty \frac{dq}{2\pi}e^{-i 2 q l} \Xi(0,L) \nonumber \\
\times e^{-i2\left(1+\frac{q}{\pi}\right)k_F^hL}  \langle e^{i2\left\{\left(1+\frac{q}{\pi}\right)\left[
\theta(0,\tau)-\theta(L,0)\right]-[\phi(0,\tau)-\phi(L,0)]\right\}} \rangle. \nonumber \\
\label{eq:Pi_bulk_gen}
\end{eqnarray}
The expression \eqref{eq:Pi_bulk_gen} is valid quite generally under the assumption that the contacts are irrelevant perturbations and do not lead to the ``end contacted" result described earlier. If we now specialize to the spin-incoherent case we have
\begin{equation}
\Xi(0,L)=\langle Z^{\vphantom{\dagger}}_{l,\vphantom{-}s}Z^{\vphantom{\dagger}}_{l,-s}
Z^{\dagger}_{0,-s}Z^{\dagger}_{0,\vphantom{-}s}\rangle=\left(\frac{1}{2}\right)^{|l|}.
\label{eq:4_Z_SI}
\end{equation}
Substituting \eqref{eq:4_Z_SI} into \eqref{eq:Pi_bulk_gen} and integrating over momentum,
\begin{eqnarray}
\Pi^{\rm{ long}}_{\rm bulk}(0,L;\tau,0) = \left(\frac{1}{2\pi \alpha_c}\right)^2 \sum_l \nonumber \\
\langle \left(\frac{1}{2}\right)^{|l|} 
\frac{1}{2}\delta[l+\frac{k_F^h}{\pi}(L+\theta(L,0)-\theta(0,\tau))]\nonumber \\
\times e^{i2\left\{\left[
\theta(0,\tau)-\theta(L,0)\right]-[\phi(0,\tau)-\phi(L,0)]\right\}}\rangle. 
\end{eqnarray}
Finally recalling that for $L \gg a$ we can replace the sum over $l$ by an integral,\cite{Fiete:prl04} we find
\begin{eqnarray}
\Pi^{\rm{ long}}_{\rm bulk}(0,L;\tau,0) = \left(\frac{1}{2\pi \alpha_c}\right)^2\frac{1}{2} e^{-\frac{k_F^hL}{\pi}\ln2}\nonumber \\
\times \langle e^{-\frac{k_F^h}{\pi}(\theta(L,0)-\theta(0,\tau))] \ln2} 
e^{i2\left\{\left[
\theta(0,\tau)-\theta(L,0)\right]-[\phi(0,\tau)-\phi(L,0)]\right\}} \rangle, \nonumber \\
\end{eqnarray}
which can be evaluated (up to phase factors) to give
\begin{equation}
\Pi^{\rm{ long}}_{\rm bulk}(0,L;\tau,0) \propto \left(\frac{1}{2\pi \alpha_c}\right)^2 e^{-\frac{k_F^hL}{\pi} \ln2} 
\left(\frac{\alpha_c^2}{L^2+v_\rho^2 \tau^2}\right)^{\gamma_{K_\rho}},
\label{eq:SILLpropgt4}
\end{equation}
where $\gamma_{K_\rho} = \frac{1}{2K_\rho}-\frac{K_\rho}{2}\left[\left(\frac{\ln 2}{\pi}\right)^2-4\right].$
The Josephson current in the wire is then
\begin{eqnarray}
\mathcal{J^{\rm long}_{\rm bulk}(\chi)} &\propto& G_1G_2 e^{-\frac{k_F^hL}{\pi} \ln2} \left(\frac{\alpha_c}{L}\right)^{2\gamma_{K_\rho}-1}\sin(\chi)\nonumber\\&\times&\int_0^\pi dx
\left(\frac{1}{1+x^2}\right)^{\gamma_{K_\rho}},\nonumber\\\label{eq:bulkJosCurntLong}
\end{eqnarray}
where we have again cut off the integral at times $\tau=L/v_\rho$.  Evidently, the critical current again scales as an exponential of the length, ${\cal J}^{\rm long}_{\rm bulk}|_{\rm critical} \propto e^{-\frac{k_F^hL}{\pi} \ln2} \left(\frac{\alpha_c}{L}\right)^{2\gamma_{K_\rho}-1}$ rather than a power law\cite{Takane:jpsj97} $\left(\frac{\alpha_c}{L}\right)^{\frac{2}{K_\rho}-1}$ in the LL regime.   This Luttinger liquid result corrects the result originally obtained by Fazio {\it et al.}\cite{Fazio:prl95,Fazio:prb96}  by taking into account the proximity effect.  Comparing the critical currents of the end contacted \eqref{eq:endJosCurntLong} case with the bulk contacted case \eqref{eq:bulkJosCurntLong}, one sees that the length dependence is not materially different: they only differ in the inconsequential power law that multiplies the exponential.

{\em End contacted SILL when $L \ll \xi$}:
To compute the Josephson current we must evaluate the single-particle Green's functions that appear in \eqref{eq:SILL-propgt-Short}.  Since $\xi$ is a property of the SC which we assume is tunnel contacted to the SILL, it is independent of the properties of the SILL.  We will further assume $a \ll L \ll \xi$.  The opposite limit of the wire length being shorter than the inter-particle spacing is not well motivated physically.  The single-particle Green's functions appearing in  \eqref{eq:SILL-propgt-Short} can readily be evaluated following the method of Appendix~\ref{app:spin_current}.  This gives
\begin{eqnarray}
G_s^{\rm end}&=&\frac{1}{2\pi \alpha_c} \left(\frac{1}{2}\right)^{k_F^h L} \langle e^{-i[\phi(\tau)-\phi(L)]}\rangle, \nonumber \\
&=&\frac{1}{2\pi \alpha_c} e^{-\frac{k_F^h L}{\pi}\ln 2}   \left(\frac{\eta}{1+e^{-\eta}e^{-v_\rho \pi \tau/L}}\right)^{\frac{1}{4K_\rho}},
\end{eqnarray}
which immediately leads to the Josephson current
\begin{eqnarray}
{\cal J}^{\rm short}_{\rm end}(\chi) \propto \Delta G_1 G_2 e^{-2\frac{k_F^hL}{\pi} \ln2} \left(\frac{L}{\alpha_c}\right)^2\sin(\chi)\nonumber \\
\times \left[\int_0^\pi dx  \left(\frac{\eta}{1+e^{-\eta}e^{-x}}\right)^{\frac{1}{4K_\rho}}\right]^2,
\label{eq:endJosCurntShort}
\end{eqnarray}
with a critical current ${\cal J}^{\rm short}_{\rm end}|_{\rm critical} \propto e^{-2\frac{k_F^hL}{\pi} \ln2} \left(\frac{L}{\alpha_c}\right)^2$.

{\em Bulk contacted SILL when $L \ll \xi$}:   For a bulk contacted SC in the short wire limit we again apply the formula \eqref{eq:SILL-propgt-Short} where the Green's functions are those appropriate for an infinite system, computed earlier in the literature in Ref.~[\onlinecite{Fiete:prl04,Fiete:rmp07}].  The result is
\begin{eqnarray}
{\cal J}^{\rm short}_{\rm bulk}(\chi) \propto \Delta G_1 G_2 e^{-2\frac{k_F^hL}{\pi} \ln2} \left(\frac{\alpha_c}{L}\right)^{2\Gamma_{K_\rho}-1}\sin(\chi)\nonumber \\
\times \left[\int_0^\pi dx  \left(\frac{1}{1+x^2}\right)^{\Gamma_{K_\rho}}\right]^2,
\label{eq:bulkJosCurntShort}
\end{eqnarray}
which evidently has the critical current ${\cal J}^{\rm short}_{\rm bulk}|_{\rm critical} \propto e^{-2\frac{k_F^hL}{\pi} \ln2} \left(\frac{\alpha_c}{L}\right)^{2\Gamma_{K_\rho}-1}$, where $\Gamma_{K_\rho}=\frac{1}{4K_\rho}-K_\rho\left[\left(\frac{\ln 2}{\pi}\right)^2-1\right]$.  

Comparing the results for bulk and end contacted, as well as short verses long wires, we find the most important result is how the Josephson critical current scales with the length of the wire.  Suppressing the relatively unimportant power law decay that multiplies the dominant exponential decay we find
\begin{eqnarray}
\mathcal{J}^{\text{long}}|_{\rm critical} &\propto& e^{-\frac{k_F^hL}{\pi} \ln2} \\
\mathcal{J}^{\text{short}}|_{\rm critical} &\propto& e^{-2\frac{k_F^hL}{\pi} \ln2} . \label{eq:JosCurntCrit}
\end{eqnarray}
so that the decay of critical current with the length of the SILL is twice as fast as for a short wire.  This originates in the fact that for a short wire electrons propagate independently so the spin incoherence affects each electron independently, rather than a single coherent pair as in the case of the long wire.
It is also worth emphasizing that the assumption of tunneling contacts always results in a Josephson current that is proportional to the product of the bare conductances of the two contacts with a sinusoidal dependence on the phase difference of the superconducting order parameter, ${\cal J}(\chi) \propto G_1G_2 \sin(\chi)$.   Therefore it is the length dependence of the critical current \eqref{eq:JosCurntCrit} that provides a smoking gun signature of spin-incoherence in the dc Josephson effect with tunnel contacts.  For Andreev (adiabatic) contacts, the length dependence is identical to those of a LL or non-interacting one dimensional system.\cite{Tilahun:prb08}  Instead, the SILL physics is manifest clearly in the flux dependence of the Josephson current which takes on a saw-tooth form of half the usual period.\cite{Tilahun:prb08}

\subsubsection{AC Josephson Effect}

The ac Josephson effect occurs when there is a finite voltage $V$ across the SC-SILL-SC system. The Josephson phase acquires a time dependence $\dot \chi = 2eV$ leading to a Josephson current that oscillates in time.  A sub-gap dissipative current  is also induced, but at
small voltages this can be estimated to be small.\cite{Fazio:prb96}  As in the case of the dc Josephson effect, the qualitative features of the ac Josephson effect also depend on whether the wire is short or long, as defined in the dc case. 

{\it Long wire case}:  We have seen that the qualitative features of the dc Josephson current  in the spin-incoherent regime for bulk and end contacts are not very different.  That is also true of the ac Josephson effect. The ac Josephson current can be computed from\cite{Fazio:prb96}
\begin{eqnarray}
\mathcal{J}(t) = 4 \pi^2 e v_F^2 \frac{G_1G_2}{(4e^2)^2} {\hskip 4cm}\nonumber \\
\times {\rm Re}\left[\sum_{\pm} \pm e^{\pm i 2eV t}\int_0^\infty dt' e^{\mp i eVt'} \Pi(t')\right],
\end{eqnarray}
where $\Pi(t)$ is the Cooper pair propagator, Eq. \eqref{eq:SILL-propgt-Long}, evaluated at real times. The ac Josephson current can be decomposed
into sinusoidal and cosinusoidal components,
\begin{equation}
\mathcal{J}(t)= 4 \pi^2 e v_F^2 \frac{G_1G_2}{(4e^2)^2}  [J_s\sin{2eVt}+J_c\cos{2eVt}],
\end{equation}
where
\begin{eqnarray}
\label{eq:J_s}
J_{s}&=&-{\rm Im}\left[\int_0^\infty dt' \left(e^{-i eVt'} \Pi(t') + e^{+i eVt'} \Pi(t')\right)\right],\;\;\;\;\\
\label{eq:J_c}
J_{c}&=&{\rm Re}\left[\int_0^\infty dt' \left(e^{-i eVt'} \Pi(t') - e^{+i eVt'} \Pi(t')\right)\right].
\end{eqnarray}
In the Luttinger liquid regime $J_s$ and $J_c$ oscillate with voltage across the junction, as does the amplitude $J_a=\sqrt{J_s^2+J_c^2}$.\cite{Fazio:prb96}  The frequency of the oscillations of $J_a$ with voltage depends on the relative spin and charge velocities.   One period occurs when the spin and charge parts differ by $2\pi$, that is when $eV=2\pi/(L/v_\sigma-L/v_\rho)$.\cite{Fazio:prb96}  In the spin-incoherent regime, we have $v_\sigma \to 0$, implying vanishing voltages  will lead to oscillations and they may cease to be observed.  The lack of amplitude oscillations will persist into the spin-incoherent regime, which has effectively only one velocity, the charge velocity.  On the other hand, if $v_\sigma$ is not too different from $v_\rho$ (say, $v_\sigma=v_\rho/10$), then one can expect to find a temperature dependence of the voltage oscillations that reveals spin-incoherent physics in a way analogous to Coulomb drag\cite{Fiete:prb06} or charge fluctuation noise:\cite{Fiete:prb07} At temperatures below the spin energy, there will be amplitude oscillations in the Josephson current as a function of voltage, while for temperatures above the spin energy there will be no such oscillations because the spin mode effectively does not propagate.  This ``washing out" of the high frequency (because of the ratio $v_\sigma/v_\rho \ll 1$) oscillations with temperature is the signature of spin-incoherent physics in the ac Josephson effect.

{\it Short wire case}:  For a short wire, the voltage dependence is independent of the properties of the 1-d system,\cite{Fazio:prb96} be it a LL or a SILL.  In this case the ac Josephson current is
\begin{equation}
{\cal J}(t) = \frac{2}{\pi} K(eV/2\Delta) {\cal J}_c(0)\sin(2eV t),
\label{eq:acJosShort} 
\end{equation}
where ${\cal J}_c(0)$ is the zero voltage critical current for the short wire given in Sec.~\ref{sec:DC-Josephson}, and $K(x)$ is an elliptical integral.  Recall that for a short wire, ${\cal J}_c(0)\propto e^{-2\frac{k_F^hL}{\pi} \ln2}$.  Thus, the ac Josephson effect is only effective at revealing spin-charge separation in the long wire limit, and there are no new spin-incoherent features that appear relative to the LL aside from the length dependence of the dc critical current that enters in \eqref{eq:acJosShort}.

\section{Discussion}
\label{sec:discussion}

In this paper we have touched on what we believe are some of the most easily observed consequences of spin-incoherent behavior in
ferromagnet/spin-incoherent Luttinger liquid and superconductor/spin-incoherent Luttinger liquid junctions.  For the case of FM-SILL junctions,
we computed the spin current pumped into the SILL as a result of magnetization dynamics and  the effect of spin accumulation in the SILL on the
parameters governing the magnetization dynamics.  We found that for interactions in the SILL with $1/2 <K_\rho <1$ the temperature dependence of
the spin current and magnetization dynamics is {\em qualitatively different} from the case of a LL and should thus be observable experimentally.
Some of the key differences between FM-SILL and FM-LL systems are summarized in Table~\ref{tb:LL_SILL}. The crossover from FM-LL to FM-SILL in
the exponent $\delta$ governing the temperature dependence of several key quantities is shown in Fig.~\ref{fig:delta}.

In the case of SC-SILL junctions our results greatly extend those obtained earlier by us.\cite{Tilahun:prb08} In that earlier work we were
concerned only with the case of adiabatic (Andreev) contact of the spin-incoherent Luttinger liquid to the superconductor.  Here we have
developed those results further and also discussed the opposite limit of tunnel contacts to the SC.  In the tunneling limit we have computed the
ac and dc Josephson response in the geometries shown in Fig~\ref{fig:SC_tunnel}.  We find that in contrast to the case of adiabatic contacts,
the tunnel contacts lead to a Josephson critical current that is exponentially suppressed with the length of the SILL region.  This difference
arises from the fact that in the Andreev limit the dc Josephson effect is determined solely by the zero modes of the Hamiltonian, while for tunnel
contacts the non-zero modes generally dominate the response.  These non-zero modes enter because in the tunneling process an electron is created
locally near the tunnel barrier and thus requires many wavevectors to build its wavepacket.

With the aim of providing a general discussion of junctions of ferromagnets or superconductors with a strongly interacting one dimensional
system, we have couched many of our calculations in the recently developed scheme for bosonizing strongly correlated electron
systems.\cite{Matveev:prb07}  This formalism is valid for arbitrary temperatures with respect to the spin energy $E_\sigma$, but requires that
the temperatures remain small compared to the charge energy $E_\rho$.   In order to adapt that formalism to the systems we discussed here, we
extended those results to express open and Andreev boundary conditions in a language valid for arbitrary temperatures with respect to the spin
sector.  Through several examples, we showed how various correlations functions could be evaluated and verified that in the spin-incoherent
regime the results properly reduce to the results obtained using the world-line picture.\cite{Fiete:prl04}

One of our primary motivations for using the language of Ref.~[\onlinecite{Matveev:prb07}] is to help provide a clear starting point for
numerical studies of the spin-incoherent regime and the many interesting (and probably experimentally relevant) crossovers that occur between it
and the Luttinger liquid regime.  Within this formalism the charge physics can be computed analytically via a standard bosonization scheme, but
the spin sector must be addressed numerically for temperatures of order the spin energy $E_\sigma$.   The types of correlation functions that
must be computed numerically are those that add and remove a site (or multiple sites) from a spin chain, such as $\langle Z_{l_1,s}
Z^\dagger_{l_2,s}\rangle$, that appear in the evaluation of a single-particle (or multi-particle) Greens function.  (See Eq.~\eqref{eq:G_app},
for example.)  We would like to emphasize that numerical studies of strongly interacting one-dimensional electron systems with appreciable
temperature compared to the spin energy is an entirely untouched area and is now ripe for investigation.

On the experimental side, the lack of a clear experimental ``smoking gun" observation of the SILL remains a key issue to be addressed. However,
there are mounting experimental indications we are not far away.\cite{Steinberg:prb06,Hew:prl08,Deshpande:np08}  Our best numerical
estimates\cite{Fiete:prb05,Fiete:rmp07} suggest that many low density quantum wires sit right on the edge of the spin-incoherent regime and
perhaps all that is needed is a focused experimental effort to search for its signatures, rather than any key technical breakthough.  One of our
aims in this work it to highlight certain classes of hybrid structures where spin-incoherent physics should be observable.

Finally, we would like to close with what we regard as some of the outstanding theoretical issues surrounding the spin-incoherent Luttinger
liquid. Perhaps the main one is the behavior on temperature scales $k_B T \approx E_\sigma$ that we already alluded to above. Related to this is
a better understanding of the crossover between the Luttinger liquid and the spin-incoherent Luttinger liquid regimes.  Both of these will
likely require a numerical attack as there are no obvious analytical methods available to address them.  There is also the issue of spin-orbit
coupling that has so far received no attention.  For very strong spin-orbit coupling is there novel behavior in the regime $E_\sigma \ll k_B T
\ll E_{SO},E_\rho$?  The subject of noise in hybrid structures involving a SILL will be discussed in a forthcoming work.\cite{Dagim_unpublished}

\acknowledgments

We thank Leon Balents for discussions on spin transport in one dimensional systems and Oleg Starykh for discussions on pair correlations in
superconductors.  This work was supported by NSF Grants PHY05-51164, DMR-0606489, the Lee A. DuBridge Foundation, and the Welch Foundation.

\appendix
\section{Evaluation of $C(\omega)$ using bosonization for strongly interacting electrons}
\label{app:spin_current} 

In this appendix we compute the correlator $C(t)$ and its Fourier transform $C(\omega)$ using the general formalism for
bosonization of strongly correlated electrons in one dimension developed by Matveev, Furusaki, and Glazman who also applied it to the evaluation
of the single particle Green's function for an infinite system and its Fourier transform.\cite{Matveev:prb07,Matveev:prl07}  We summarized the
main results of the bosonization scheme in Sec.~\ref{sec:bosonization}.    The calculation here is for a semi-infinite, or finite system with
open boundary conditions, so is different in detail from what has been discussed in Refs.~[\onlinecite{Matveev:prb07,Matveev:prl07}], but the
basic elements of the bosonization are the same.  This appendix is meant to illustrate clearly in a specific example which parts of correlation
functions can be computed analytically at finite temperatures and which pieces in general require methods yet to be developed, or a numerical
attack.  As was already discussed in Sec.~\ref{sec:bosonization} the chief difficulty lies in computing the correlations at arbitrary
temperatures in the spin sector.  It is hoped that the details given here will provide a good starting point for those skilled in numerics to
enter the study of strongly interacting one dimensional systems where there is currently {\em no} quantitative understanding of the regime $k_B
T \approx E_\sigma\ll E_\rho$.  Given the typical values of $T,E_\sigma,E_\rho$
present\cite{Auslaender:sci05,Steinberg:prb06,Fiete:prb05,Hew:prl08,Halperin:jap07} in quantum wires, this ``intermediate" temperature regime
may turn out to be the most relevant experimentally.

In the evaluation of the spin current \eqref{spnexpr} (pumped from a ferromagnet into an adjoined one dimensional system coupled via a tunnel
junction), the correlation function
\begin{equation}
 C(t) =-i\Theta(t)\sum_s\langle[F^\dagger_s(t)\psi_s(t),\psi^\dagger_s(0)F_s(0)]\rangle
 \end{equation}
arises in lowest (second) order perturbation theory. It contains the difference of the product of single particle Greens functions for both the
FM and the interacting one dimensional system
 \begin{eqnarray}
 C(t) =-i\Theta(t)\sum_s[\langle F^\dagger_s(t)F_s(0)\rangle \langle \psi_s(t)\psi^\dagger_s(0)\rangle\nonumber \\
 -\langle F_s(0)F^\dagger_s(t)\rangle \langle \psi^\dagger_s(0) \psi_s(t)\rangle],
 \end{eqnarray}
where the brackets $\langle \cdot \rangle$ denote a thermal average computed with the open boundary conditions described in
Sec.~\ref{subsec:open}. The correlators $\langle F^\dagger_s(t)F_s(0)\rangle$ and $\langle F_s(0)F^\dagger_s(t)\rangle$ can be computed by
standard bosonization methods.\cite{Giamarchi}  The result is
\begin{equation}
\label{eq:FF_t}
\langle F^\dagger_s(t)F_s(0)\rangle=i\langle F_s(0)F^\dagger_s(t)\rangle=
\frac{i}{2\pi \alpha_c}\frac{\left(\pi k_B T/\epsilon_c\right)}{\sinh\left( \frac{\pi k_B T t}{\hbar}\right)},
\end{equation}
where $\epsilon_c=\hbar v_F/\alpha_c$ is a high energy cut off.

Our real objects of interest here are  the boundary Greens functions $G^+_s(t)\equiv \langle \psi_s(t)\psi^\dagger_s(0)\rangle$ and
$G^-_s(t)\equiv \langle \psi^\dagger_s(0) \psi_s(t)\rangle$ evaluated at arbitrary temperature with respect to $E_\sigma$, but small compared to
$E_\rho$.  To evaluate these we make use of expression \eqref{eq:psi_final_bos} for the electron operator. Straight forward substitution gives
\begin{eqnarray}
\label{eq:G_app}
G^+_s(t)=\frac{1}{2\pi \alpha_c} \int_{-\infty}^{\infty} \frac{dq_1}{2\pi} \int_{-\infty}^{\infty} \frac{dq_2}{2\pi}
\sum_{l_1,l_2} e^{-i(q_1 l_1-q_2l_2)} \nonumber \\
\times \langle e^{i[(1+\frac{q_1}{\pi})\theta(t)-\phi(t)]} Z_{l_1,s} Z^\dagger_{l_2,s} e^{-i[(1+\frac{q_2}{\pi})\theta(0)-\phi(0)]}\rangle.
\end{eqnarray}
Open boundary conditions at $x=0$ prevent fluctuations of the $\theta$ field which sets it to a constant, which we take to be 0. With $\theta$
set to a constant, the integrals over momentum can be done trivially resulting in $\delta$-functions for $l_1,l_2$ which kill those sums and
selects $l_1=l_2=0$.  The final result is remarkably simple
\begin{equation}
  G^+_s(t)= \frac{1}{2\pi \alpha_c}\langle Z_{0,s}(t) Z^\dagger_{0,s}(0)\rangle \langle  e^{-i[\phi(t)-\phi(0)]}\rangle.
\end{equation}
Similar manipulations yield
\begin{equation}
  G^-_s(t)= \frac{i}{2\pi \alpha_c}\langle Z^\dagger_{0,s}(0) Z_{0,s}(t)\rangle \langle  e^{-i[\phi(t)-\phi(0)]}\rangle,
\end{equation}
where we have made explicit the time dependence in the spin correlators involving $Z^\dagger_{0,s}$ and $Z_{0,s}$. The general theoretical
challenge is then to compute those correlation functions which involve adding and removing a site from the spin chain, which is an interacting
many-body problem analogous to Fermi edge physics\cite{Fiete:prl06,Ogawa:prl92,Kane:prb94} only it involves knowledge of the Hamiltonian on
potentially many energy scales.  However, for $E_\sigma \ll E_\rho$ the time evolution of the spin degrees of freedom are very slow compared to
the charge degrees of freedom, regardless of temperature, and can be neglected for times $\lesssim \hbar/E_\sigma$. Note that in the limit
$E_\sigma \to 0$, the spin dynamics can always be neglected.  Furthermore, if $k_B T \gg E_\sigma$, then the spins become essentially
non-interacting, and therefore non-dynamical. Then the correlation functions simplify considerably. In the spin-incoherent regime one
has\cite{Matveev:prl07,Fiete:prl04,Penc_Serhan:prb97}
\begin{equation}
\label{eq:Z+Z}
\langle Z^\dagger_{0,s}Z_{0,s}\rangle={1 \over 2},
\end{equation}
and
\begin{equation}
\label{eq:ZZ+}
\langle Z_{0,s} Z^\dagger_{0,s}\rangle=1,
\end{equation}
independent of time which can be straightforwardly generalized to include an applied external magnetic
field.\cite{Fiete:prb05,Fiete:prl06,Kindermann_crossover:prb06,Kindermann:prl06}  The result is
\begin{eqnarray}
\langle Z^\dagger_{0,s}Z_{0,s}\rangle&=& p_s,    \\
\langle Z_{0,s} Z^\dagger_{0,s}\rangle&=& 1,
\end{eqnarray}
where $p_s$ is the probability of having spin projection $s$.  It takes the values $p_\uparrow=1-p_\downarrow=\frac{1}{e^{-E_Z/k_BT}+1}$, where
$E_Z$ is the Zeeman energy.  Note that the correlation functions $\langle Z^\dagger_{0,s}Z_{0,s}\rangle$ and $\langle Z_{0,s}
Z^\dagger_{0,s}\rangle$  in the spin-incoherent regime are identical to their values in the infinite system.\cite{Matveev:prb04}  This is
because in the spin-incoherent regime the spins are non-dynamical, so the open boundary conditions which tend to suppress fluctuations have
essentially no effect on the spins which are rendered non-dynamical by $k_B T \gg E_\sigma$.

Finally, we are left to evaluate $ \langle  e^{-i[\phi(t)-\phi(0)]}\rangle$, which is identical to $\langle F^\dagger_s(t)F_s(0)\rangle$ and
$\langle F_s(0)F^\dagger_s(t)\rangle$ only with the exponent changed
\begin{equation}
\label{eq:psipsi_t} \langle  e^{-i[\phi(t)-\phi(0)]}\rangle =\frac{(-i)^{1/2K_\rho}}{2\pi \alpha_c}\left[\frac{\left(\pi k_B
T/\epsilon'_c\right)}{\sinh\left( \frac{\pi k_B T t}{\hbar}\right)}\right]^{1/2K_\rho},
 \end{equation}
where $\epsilon'_c=\hbar v_\rho/\alpha_c$ is a high energy cut off for the charge sector in the SILL.

Combining the results \eqref{eq:FF_t}, \eqref{eq:Z+Z},  \eqref{eq:ZZ+}, and \eqref{eq:psipsi_t}, we find
\begin{equation}
\label{eq:C_t_careful} C(t)=-i\Theta(t)(-i)^{1\over 2K_\rho} \left(\frac{1}{2\pi \alpha_c}\right)^2 \left[\frac{\left(\pi k_B
T/\tilde\epsilon_c\right)}{\sinh\left(\frac{\pi k_B T t}{\hbar}\right)}\right]^{\delta^{SILL}+2},
\end{equation}
where $\delta^{SILL}=1/2K_\rho -1$ is given in Table~\ref{tb:LL_SILL}, and $\tilde \epsilon_c$ is an effective high energy cut off given by
$\tilde\epsilon_c=\left[\epsilon_c (\epsilon'_c)^{1/2K_\rho}\right]^{2K_\rho/(2K_\rho+1)}$. For making the following formulas more compact, we
define $\tilde \delta \equiv \delta^{SILL}+2$.

We are now ready to compute the Fourier transform of \eqref{eq:C_t_careful}, $C(\omega)=\int_{-\infty}^\infty dt e^{i\omega t} C(t)$. Making the
change of variables $X=\frac{\pi k_BT}{\hbar} t$, we have
\begin{equation}
C(\omega) =\frac{ (-i)^{1+\frac{1}{ 2K_\rho}}}{(2\pi \alpha_c)^2} \left(\frac{\hbar}{\pi k_B T}\right) \left(\frac{\pi k_B T}{\tilde
\epsilon_c}\right)^{\tilde \delta}\int_0^\infty dX \frac{e^{i \left(\frac{\hbar \omega}{\pi k_B T}\right) X}}{[\sinh(X)]^{\tilde \delta}}.
\end{equation}
The integral is standard
\begin{equation}
\int_0^\infty dX \frac{e^{i \left(\frac{\hbar \omega}{\pi k_B T}\right) X}}{[\sinh(X)]^{\tilde \delta}} =2^{\tilde \delta-1}\Gamma(1-\tilde
\delta)\frac{\Gamma\left( {\tilde \delta \over 2}-i\frac{\hbar \omega}{2\pi k_B T}\right)}{\Gamma\left(1-{\tilde \delta \over 2}-i\frac{\hbar
\omega}{2\pi k_B T}\right)}.
\end{equation}
The result of the integral can be transformed to a more convenient form using the following identities for complex number $z$: $\pi=\sin(\pi
z)\Gamma(z)\Gamma(1-z)$ and $\Gamma(z)^*=\Gamma(z^*)$. This gives
\begin{eqnarray}
\frac{\Gamma\left( {\tilde \delta \over 2}-i\frac{\hbar \omega}{2\pi k_B T}\right)}{\Gamma\left(1- {\tilde \delta \over 2}-i\frac{\hbar \omega}
{2\pi k_B T}\right)}=\Bigg|\Gamma\left({\tilde \delta \over 2}+i\frac{\hbar \omega}{2\pi k_B T}\right)\Bigg|^2 \nonumber \\
\times \sin\left[\pi\left({\tilde \delta \over 2}+i\frac{\hbar \omega}{2\pi k_B T}\right)\right] {1 \over \pi},
\end{eqnarray}
where the sin can be expanded to pick out the real and imaginary parts.  Selecting the imaginary part of $C(\omega)$ we find
\begin{eqnarray}
{\rm Im}[C(\omega)]\propto \frac{1}{(2\pi \alpha_c)^2}\left(\frac{\hbar}{\pi k_B T}\right)
\left(\frac{\pi k_B T}{\tilde \epsilon_c}\right)^{\tilde\delta}\nonumber \\
\times \sinh\left(\frac{\hbar \omega }{2\pi k_B T}\right)\Bigg|\Gamma\left({\tilde \delta \over 2}+i\frac{\hbar \omega}{2\pi k_B T}\right)\Bigg|^2,
\end{eqnarray}
which for $\hbar \omega \ll k_B T$ gives
\begin{equation}
{\rm Im}[C(\omega)]\propto \hbar \omega (k_B T)^{\delta^{SILL}},
\end{equation}
in agreement with our \eqref{eq:C_omega_sketch}, and the general analytical form earlier obtained by Bena and Balents\cite{Bena:prb04} up to the
exponent describing the temperature dependence which is changed in the spin-incoherent regime, as illustrated in Fig~\ref{fig:delta}.

The calculation above is valid provided the temperature is larger than the level spacing of the finite length wire: $k_B T \gtrsim \hbar
v_\rho/L$. For $k_B T \lesssim \hbar v_\rho/L$ the finite length results of Ref.~[\onlinecite{Kane:prl97}] can readily be generalized to finite
temperatures. No simple power laws emerge, but rather a more complicated dependence involving hyperbolic trig functions.

\section{SC-SILL junctions in the Andreev limit}
\label{app:Andreev_TDOS}

In this appendix we show that the bosonization scheme introduced in Sec.~\ref{sec:bosonization} recovers our earlier results for the
single-particle Greens function, the tunneling density of states, and the pair correlations.\cite{Tilahun:prb08}

\subsection{Single-particle Greens function}

We are interested in computing the Fourier transform of the single-particle Green's function $G^+_s(x,x',t)=\langle
\psi_s(x,t)\psi_s^\dagger(x',0)\rangle$ where here $x$ and $x'$ measure the distance from the boundary with the SC.  Taking $x=x'$ and using the
general formula \eqref{eq:psi_final_bos} we have
\begin{eqnarray}
G^+_s(x,t)=\frac{1}{2\pi \alpha_c} \int_{-\infty}^{\infty} \frac{dq_1}{2\pi} \int_{-\infty}^{\infty} \frac{dq_2}{2\pi}
\sum_{l_1,l_2} e^{-i(q_1 l_1-q_2l_2)}  \nonumber \\
\times \langle Z_{l_1,s} Z^\dagger_{l_2,s} \rangle \langle e^{i[(1+\frac{q_1}{\pi})[k_F^hx+\theta(x,t)]-\phi(x,t)]}\nonumber \\
\times  e^{-i[(1+\frac{q_2}{\pi})[k_F^hx+\theta(x,0)]-\phi(x,0)]}\rangle,\;\;\;
\end{eqnarray}
where the $\theta$ and $\phi$ fields have the expansion given in \eqref{eq:Andreev_expansions}. We take the limit $L\to \infty$ as we did in
Ref.~[\onlinecite{Tilahun:prb08}] so that the zero modes play no role. Since the holon sector is described the the Gaussian theory
\eqref{eq:H_holon} we can make use of the identity $\langle e^{i A}\rangle= e^{-\langle A^2\rangle/2}$ for operator $A$ and evaluate the
correlation functions at zero temperature with respect to the charge energy but infinite temperature with respect to the spin energy: $E_\rho
\gg  k_B T\to 0 \gg E_\sigma \to 0$, that is we take both the spin energy and the temperature to zero but alway maintain $k_B T \gg
E_\sigma$.\cite{Fiete:prl04}  In the spin-incoherent regime, the spin correlations are translationally invariant so we have $ \langle Z_{l_1,s}
Z^\dagger_{l_2,s}\rangle= \langle Z_{l_1-l_2,s} Z^\dagger_{0,s}\rangle=1/2^{|l_1-l_2|}$. Making a change of variables $l=l_1-l_2$, the summation
over the sites of the spin chain results in a delta function, $2\pi \delta(q_1-q_2)$, which immediately kills one of the momentum integrals and
sets $q_1=q_2$.  We re-label the remaining momentum variable $q$,
\begin{eqnarray}
G^+_s(x,t)=\frac{1}{2\pi \alpha_c} \int_{-\infty}^{\infty} \frac{dq}{2\pi}\sum_{l=-\infty}^\infty \langle 2^{-|l|} e^{-i ql}
\nonumber \\ \times  e^{i[(1+\frac{q}{\pi})\theta(x,t)-\phi(x,t)]}  e^{-i[(1+\frac{q}{\pi})\theta(x,0)-\phi(x,0)]}\rangle.\;\;\;
\end{eqnarray}
Next, we define a variable $Y=1+q/\pi$ and re-express the integration in terms of this variable,
\begin{eqnarray}
\label{eq:G_final_trans}
G^+_s(x,t)=\frac{1}{2 \pi \alpha_c} \int_{-\infty}^{\infty} \frac{dY}{2}\sum_{l=-\infty}^\infty  \langle 2^{-|l|} (-1)^{l} e^{-i l \pi Y}
\nonumber \\ \times  e^{i[Y\theta(x,t)-\phi(x,t)]}
 e^{-i[Y\theta(x,0)-\phi(x,0)]}\rangle.
\end{eqnarray}
Finally, noting that $\delta(\pi l- [\theta(x,t)-\theta(x,0)])=\int _{-\infty}^{\infty} \frac{dY}{2\pi} e^{-i[\pi l -
(\theta(x,t)-\theta(x,0))]Y}$ we see that Eq.~\eqref{eq:G_final_trans} correctly reproduces Eq.(4) of Ref.~[\onlinecite{Tilahun:prb08}] after we
recall the relations $ \phi_\rho/\sqrt{2} =\phi$ and $\sqrt{2 }\theta_\rho=\theta$.\cite{Fiete_2:prb05,Fiete:rmp07}  The identical result
therefore follows for the Greens function and the tunneling density of states derived from it.

\subsection{Pair correlations}

Here we compute the pair correlation, $F(x)=-\langle \psi_\uparrow(x) \psi_\downarrow(x)\rangle$, a distance $x$ from the superconductor
boundary. By direct substitution of \eqref{eq:psi_final_bos} and \eqref{eq:little_z} we have
\begin{eqnarray}
F(x)=-\frac{1}{2\pi \alpha_c} \int_{-\infty}^{\infty} \frac{dq_1}{2\pi} \int_{-\infty}^{\infty} \frac{dq_2}{2\pi}
\sum_{l_1,l_2} e^{-i(q_1 l_1-q_2l_2)}  \nonumber \\
\times
\langle e^{-i\phi(x)} Z_{l_1,\uparrow} e^{i[(1+\frac{q_1}{\pi})[k_F^hx+\theta(x)]-\phi(x)]}\nonumber \\
\times e^{-i\phi(x)} Z_{l_2,\downarrow} e^{i[(1+\frac{q_2}{\pi})[k_F^hx+\theta(x)]-\phi(x)]}\rangle.
\end{eqnarray}
At the boundary, we showed in Sec.~\ref{subsec:Andreev} that Andreev reflection implies that $Z_{l,\uparrow}=Z^\dagger_{l,\downarrow}$. Since
$l=\theta(x=0)$ at the boundary we may fluctuate, we expect also that the result $Z_{l,\uparrow}=Z^\dagger_{l,\downarrow}$ approximately holds
near (within $\xi$) of the boundary. Hence,
 \begin{eqnarray}
F(x)\approx-\frac{1}{2\pi \alpha_c} \int_{-\infty}^{\infty} \frac{dq_1}{2\pi} \int_{-\infty}^{\infty} \frac{dq_2}{2\pi}
\sum_{l_1,l_2} e^{-i(q_1 l_1-q_2l_2)}  \nonumber \\
\times
\langle e^{-i\phi(x)} Z_{l_1,\uparrow} e^{i[(1+\frac{q_1}{\pi})[k_F^hx+\theta(x)]}\nonumber \\
\times e^{-i\phi(x)} Z^\dagger_{l_2,\uparrow} e^{i[(1+\frac{q_2}{\pi})[k_F^hx+\theta(x)]}\rangle.
\end{eqnarray}
We now assume that we are in the spin-incoherent regime where the spin correlations are translationally invariant: $\langle Z_{l_1,s}
Z^\dagger_{l_2,s}\rangle= \langle Z_{l_1-l_2,s} Z^\dagger_{0,s}\rangle=1/2^{|l_1-l_2|}$.  Then repeating the steps used above to compute the
single particle Greens function (changing the summation variable $l=l_1-l_2$, killing the momentum integral with $\delta$-function), we find
 \begin{eqnarray}
F(x)\approx-\frac{1}{2\pi \alpha_c} \int_{-\infty}^{\infty} \frac{dq}{2\pi}
\sum_{l} \langle e^{-iq l}2^{-|l|}  \nonumber \\
\times   e^{i2\frac{q}{\pi}[k_F^hx+\theta(x)]} e^{-i2\phi(x)} \rangle.
\end{eqnarray}
We then make use of the integral representation of the $\delta$-function, $  \int_{-\infty}^{\infty} \frac{dq}{2\pi}\sum_{l}
e^{-iq\left(l-\frac{2}{\pi}[k_F^hx+\theta(x)]\right)} =\delta \left(l-\frac{2}{\pi}[k_F^hx+\theta(x)]\right)$ to obtain
\begin{equation}
F(x)\approx-\frac{1}{2\pi \alpha_c}\sum_l \langle 2^{-|l|}\delta\left (l-\frac{2}{\pi}[k_F^hx+\theta(x)]\right)e^{-i2\phi(x)} \rangle,
\end{equation}
which for $x \gg a,\alpha_c$ can be evaluated by taking the discrete sum to an integral: $\sum_l \to \int dl$. Doing so we recover the result
(9) of Ref.~[\onlinecite{Tilahun:prb08}] after again recalling the relations $ \phi_\rho/\sqrt{2} =\phi$ and $\sqrt{2
}\theta_\rho=\theta$.\cite{Fiete_2:prb05,Fiete:rmp07}


\begin{thebibliography}{77}
\bibitem[{Tak()}]{Takane_remark}
\bibinfo{note}{While our expansions differ slightly from the expansions in
  Ref.~[\onlinecite{Takane:jpsj96}], they give rise to the same commutation
  relations and correlation functions once our differences in zero mode signs
  are also taken into account. The difference between the two is in the
  placement of minus signs--some overall factors and some factors separating
  the $b_m, b^\dagger_m$ with the necessary $i$'s to keep the field expansion
  Hermitian. We also remark that while the field expansions look to be only a
  function of $x+\xi/2$, one can see by expanding near $x=L$, that the fields
  could also be written as $x-\xi/2$ (up to an unimportant overall minus sign
  that drops out of the correlation functions and commutation relations)
  preserving the symmetry of the proximity effect near the $x=0$ boundary and
  the $x=L$ boundary.}
\bibitem[{coh()}]{coherence_length_remark}
\bibinfo{note}{Compared to the LL case discussed in this language
  earlier\cite{Fazio:prb96,Fazio:prl95} which implicitly assumed that the spin
  velocity $v_\sigma$ is not much different from the charge velocity $v_\rho$
  (and moreover that the Fermi velocity is the same in SC and LL), here there
  is an additional length scale due to the possibility that the spin and charge
  velocities can be exponentially separated.\cite{Fiete:rmp07} In the case of
  Andreev boundary conditions we have not investigated in detail whether the
  superconducting proximity length near the boundary is necessarily set by the
  inter-particle spacing $a$, or whether it could be parametrically larger than
  $a$. For the case of tunnel junctions computed in the perturbative limit that
  we study here, the issue of different length scales does not arise.}
\end{thebibliography}

\end{document}